\documentclass[12pt]{article}
\usepackage{amsmath, latexsym, amssymb, amscd, slashed}

\numberwithin{equation}{section}
\newcommand{\R}{\mathbb R}
\newcommand{\C}{\mathbb C}

\newcommand{\Bold}[1]{{\boldsymbol{\mathit{#1}}}}

\newcommand{\Z}{\mathrm{Z}}

\newcommand{\QED}{\hspace{.2in}\square\newline}

\newtheorem{example}{Example}[section]

\newtheorem{corollary}{Corollary}[section]
\newtheorem{proposition}{Proposition}[section]
\newtheorem{definition}{Definition}[section]
\newtheorem{assumption}{Assumption}[section]

\newtheorem{remark}{Remark}[section]
\begin{document}

\title{Functional Integral Approach to $C^*$-algebraic\\ Quantum Mechanics I:\\ Heisenberg and Poincar\'{e}}
\author{J. LaChapelle}

\maketitle

\begin{abstract}
The algebraic approach to quantum mechanics has been vital to the development of  quantum theory since its inception, and it has evolved into a mathematically rigorous $C^\ast$-algebraic formulation of the theory's axioms. Conversely, the functional approach in the form of Feynman path integrals is far from mathematically rigorous: Nevertheless, path integrals provide an equally valid and useful formulation of the axioms of quantum mechanics. The two approaches can be merged by employing a notion of functional integration based on topological groups that allows to construct functional integral representations of $C^\ast$-algebras. The merger achieves a hybrid formulation of the axioms of quantum mechanics in which topological groups play a leading role. To illustrate the formalism, we apply the framework to non-relativistic and relativistic quantum mechanics via the Heisenberg and Poincar\'{e} groups.
\end{abstract}

\section{Introduction}
The axioms of quantum mechanics (QM) can be roughly classified as kinematical and dynamical. On the kinematical side sit the notions of Hilbert space of states, self-adjoint operators, and the Born rule; while on the dynamical side sit the unitary evolution operator, the evolution equation, and the outcome of observation/measurement. The axioms  themselves are typically realized using either of two complementary strategies; the functional approach as embodied in the Feynman path integral or the $C^\ast$-algebraic approach.\footnote{Our comments are restricted to QM where the functional and algebraic approaches are well understood and enjoy equal status. Unfortunately, the algebraic approach to quantum field theory (QFT) for interacting fields is notoriously problematic. On the other hand, the functional integral approach, with its S-matrix interpretation, is quite developed and successful and consequently dominates in QFT. We do not address QFT in this paper; except briefly in \S\ref{Poincare}.}

The algebraic realization of the axioms was initiated by von Neumann\cite{VN} and extended in \cite{GN},\cite{SE}.  It effectively culminated in the Gelfand-Naimark-Segal construction and the Gelfand-Naimark theorem allowing the kinematical QM axioms to be formulated in terms of certain linear functionals on $C^\ast$-algebras. By now the structure has been vastly elaborated and we refer to \cite{KR}, \cite{BEH} for details and more references. The upshot is there exists a powerful and well-developed $C^\ast$-algebraic framework on which to base the kinematics of QM.

But the dynamics --- which includes both evolution and observation --- requires further input and interpretation: Evolution is naturally formulated by specifying a one-parameter group of $\ast$-automorphisms of the algebra, while the observation/measurement issue remains unresolved (or at least controversial) and open to interpretation. Dynamical evolution leads to the study of derivations on the algebra; and eventually, correspondence with classical physics fixes the nature of the derivations --- though, famously, not uniquely in general. This somewhat awkward inclusion of dynamics in the algebraic formulation has stimulated further work to develop a single encompassing structure.

One approach to encompass both the kinematics and dynamics of QM under one algebraic roof utilizes \emph{crossed products} (which are reviewed from a physics perspective in \cite{LANDS1}--\cite{LANDS3} and reviewed and rigorously developed in \cite{W}). Crossed products were introduced in \cite{DKR} and are closely allied with the work of Mackey\cite{M1}--\cite{M4} on representation theory.

The relevance and utility of crossed products in the  formulation of QM is: i) they provide a single algebraic structure built from the original $C^\ast$-algebra and its $\ast$-automorphism group that encodes kinematics and dynamics, and ii) they realize  $\ast$-representations of the integrated Heisenberg equation on an associated Hilbert space. Stated precisely, there is a one-to-one correspondence between a covariant representation of a dynamical system (which, in particular, encodes the integrated Heisenberg equation of motion) and non-degenerate representations of the crossed product (\cite{W}\,\S2.2--2.4). In short, crossed products provide a convenient implementation of the $C^\ast$-algebraic approach to QM --- both kinematics and dynamics.

On the other side of the divide, the Feynman path integral\cite{F}--\cite{K} lies at the heart of the functional approach. The application of path integrals in standard QM is well understood, and they simultaneously incorporate both kinematics and dynamics. However, being formal objects, path integrals are often difficult to apply in more general settings with surety. Nevertheless, it is difficult to overstate the value of the physical intuition inherent in this approach.

Of course there exists already a bridge between the functional and algebraic approaches; but, other than supplying a translation dictionary, it is of limited use. This is unfortunate because it effectively separates the formal/heuristic appeal of path integrals and the rigorous mathematical development of $C^\ast$-algebras.

The purpose of this paper is to describe and illustrate what can be characterized as a hybrid of both approaches. The hybrid formulation combines the functional integral and crossed product formulations, and it is determined principally by an underlying topological group along with associated unitary representations. The rationale for introducing functional integrals into the algebraic approach should be obvious: Formal and/or heuristic application of path integrals in the functional approach is quite useful, and one hopes to enjoy similar benefits by applying functional integral techniques in the $C^\ast$-algebra setting.

Our main tool, the functional Mellin transform\cite{LA2}, provides a generalization of the standard Feynman path integral. Under suitable conditions, \emph{the space of Mellin integrable functionals is a $C^\ast$-algebra, and the functional Mellin transform implements a $\ast$-representation.} In particular, the functional Mellin transform allows to define and represent quantum observables, their traces, their logs, and their determinants as functional integrals.

Since the functional Mellin transform and its associated algebra of integrable functionals is closely related to crossed products, our approach to quantization is likewise closely related to crossed-product-inspired quantization. Indeed, mathematically, quantizations based on the two different approaches may overlap or even coincide; depending on the particular system under consideration. However, there is an important conceptual difference.

The latter employs (reduced) crossed products of $C^\ast$-dynamical systems to model the quantum algebra. At the heart of this approach is a $G$-space $(G,X)$ comprised of a topological space $X$ (usually interpreted as a classical configuration/phase space) and a locally compact transformation group $G$ that acts on $X$. One then constructs a \emph{non-commutative} dynamical system from the algebra $C_0(X)$ (continuous, bounded complex-valued functions on $X$ that vanish at infinity)  and $G$-induced homomorphisms $\alpha:G\rightarrow\mathrm{Aut}(C_0(X))$. The salient feature is the starting point of the construction, viz. $X$ and the \emph{commutative} algebra $C_0(X)$ --- i.e. crossed-product quantization views quantization as a promotion of a classical system. This of course is consistent with the time-honored strategy: $\mathrm{classical}\rightarrow\mathrm{quantum}$.

Meanwhile, we use functional Mellin transforms based on a topological group $G$ to construct a $C^\ast$-algebra (assumed to model the quantum algebra) from Mellin-integrable, operator-valued functions. This tack allows to focus on the guiding principle of symmetry without relying on an assumed classical structure, but it asks much more of $G$ than the crossed-product approach. For us, $G$ is a topological group revealed through measurement/observation as a homomorphic family of locally compact topological groups (as opposed to a transformation group on $X$). The topological group spawns the quantum Hilbert space, the $C^\ast$-algebra, and the inner-automorphisms that generate dynamics. Any classical configuration/phase space must come from the spectrum of the group's functional-Mellin-associated algebra and, in this sense, is emergent (or more properly, expected/observed). Here, the strategy is reversed: $\mathrm{quantum}\rightarrow\mathrm{classical}$.

To demonstrate our approach, we apply the formalism to the Heisenberg and Poincar\'{e} groups; reaching Poincar\'{e} by group contraction of $SO(5,\C)$. Quantizing based on the Heisenberg group, not surprisingly, is essentially equivalent to the crossed-product approach and readily leads to non-relativistic QM. On the other hand, the Poincar\'{e} group leads to a relativistic theory, with Lorentz-invariant transition amplitudes, that displays certain aspects of a QFT: But it is certainly not a quantum field theory.  Finally, we briefly consider group contraction of the Langlands dual $Sp(4,\C)$ and uncover a connection to the spinor-helicity description of scattering. Consequently, in this case Langland duals encode a dual spinor/vector parametrization of physical states. Contemplating the supergroup $OSp(5|4)$, a unified spinor/vector description of dynamics obtains in which spacetime is a secondary notion and the SUSY pairing of states reflects an equivalence between the notions of physical state scattering and physical state propagation. This has the flavor of particle/wave duality. We mention these applications are only sketched here since our ultimate goal is to present, in a companion paper, a detailed functional Mellin quantization based on $Sp(8,\R)$  (which we suspect supersedes Poincar\'{e}).

\section{Quantization}
The two primary ingredients required to implement algebraic quantization are a $C^\ast$-algebra and a group of $\ast$-automorphisms of said algebra. Incorporating functional integrals into the picture pays immediate dividends: it strongly suggests that a single object--- a topological group --- generates the entire structure. In consequence, given a (generally non-abelian) topological group,  the functional integral framework to which we adhere (see appx. A) provides a $C^\ast$-algebra of suitable functionals along with its associated inner $\ast$-automorphisms. The natural idea is that this structure models both the kinematics and dynamics of a closed quantum system.

\subsection{Preliminaries}\label{preliminaries}
Let's set the stage for quantization. The kinematic input is: (i) a $C^\ast$-algebra $\mathfrak{A}_L$ \emph{equipped with a Lie bracket structure} whose self-adjoint elements encode observable properties of some  quantum system, (ii) an associated Hilbert space of state vectors $\mathcal{H}$ that furnishes a physically relevant representation $\pi:\mathfrak{A}_L\rightarrow L_B(\mathcal{H})$ where $L_B(\mathcal{H})$ is the algebra of bounded linear operators on $\mathcal{H}$, and (iii) suitable linear functionals to account for the Born rule.

Now, a particularly interesting subset of elements of the algebra $\mathfrak{A}_L$ is its set of units, i.e. invertible elements.  Let $A_L$ be the group (or a subgroup) of units of $\mathfrak{A}_L$ and let $G$ denote a topological group isomorphic to $A_L$. By definition, $G$ is a topological linear Lie group since $A_L$ is endowed with a Lie bracket.\cite[def. 5.32]{HM} Construct the complexified group $G^\C$. The plan is to model $\mathfrak{A}_L$ by a certain $C^\ast$-algebra of functionals\footnote{We use the term functional to refer to an operator-valued function on some topological space. Strictly, functional refers to a scalar-valued function on some (usually infinite-dimensional) space. However, in the context where operator-valued functions on topological spaces are integrands of functional integrals, we will continue to use the term imprecisely. The algebra of functionals will be constructed below with the help of functional Mellin transforms.} $\mathrm{F}:G^\C\rightarrow L_B(\mathcal{H})$. Of course this algebra is not likely to be equivalent to $\mathfrak{A}_L$. But in practice one  doesn't know $\mathfrak{A}_L$ explicitly anyway: Invariably, one starts with some symmetries that characterize a system, identifies an associated group (which is typically a Lie group), and attemps to construct $\mathfrak{A}_L$ from there. So we might as well make this assumption:
 \begin{assumption}\label{first assumption}
 The $C^\ast$-algebra that characterizes a quantum system can be modeled by a certain subspace of integrable (details follow) functionals $\mathbf{F}(G^\C)\ni\mathrm{F}:G^\C\rightarrow L_B(\mathcal{H})$ where $G^\C$ is a topological linear Lie group whose Lie algebra possesses a triangular decomposition\footnote{We add this qualifier as it will simplify the discussion of representations. But more importantly the decomposition characterizes physically relevant quantum numbers via the adjoint representation. A triangular decomposition is not significantly restrictive from a physics perspective, because it includes all finite-dimensional and Kac-Moody Lie algebras. Moreover, algebras of inhomogeneous groups can be obtained by contraction.} and the Hilbert space $\mathcal{H}$ is constructed from suitable (sub)-representations of $G^\C$.
 \end{assumption}

 According to the assumption, quantization (partly) corresponds to identifying a topological linear Lie group, constructing its relevant representations, and building a suitable space of integrable functionals. But generically, $G^\C$ is topologically non-compact so it is not possible to directly extract measurable (in the mathematical sense) objects. So the assumption covers the first two points of kinematic input but does not address the Born rule. Fortunately, measure \emph{is} possible with \emph{locally compact} topological groups, and this underlies our entire construction/definition/notion of functional integrals.\cite{LA3} We need, then, some rationale to obtain locally compact topological groups from $G^\C$.

 Let $G^\C_{\Lambda}:=\{G^\C_{\lambda},\lambda\in\Lambda\}$ represent a \emph{countable} family of \emph{locally compact} topological Lie groups $G^\C_{\lambda}$ indexed by surjective homomorphisms $\lambda:G^\C\rightarrow G^\C_{\lambda}$. Since $\lambda$ is surjective, one can view $G^\C_{\lambda}$ as a subset $G^\C_{\lambda}\subset G^\C$ and interpret $\lambda$ as a `topological localization'. We will be purposely nonspecific about the set $\Lambda$, because it depends on the particular quantum system under consideration. But in general it represents constraints, state preparation/observation, and any other system particulars that an observer must specify to implement the Born rule --- which requires the structure of a measure. The point is, $G^\C$ inherits a Lie bracket structure from the quantum $C^\ast$-algebra  of a particular quantum system that can only be glimpsed as a member of $G^\C_{\Lambda}$ through interaction with or observation by an external agent.

 This leads to the second assumption:
 \begin{assumption}
A query\footnote{By `query' we mean any observation one may perform that leads to a measurable (in the operational sense) quantity.} of a quantum system corresponds to a surjective homomorphism $\lambda: G^\C\rightarrow G_{\lambda}^\C$ where $G_{\lambda}^\C$ is a \textbf{locally compact} topological linear Lie group.
 \end{assumption}

 \begin{example}
 A good example of `topological localization' is the familiar Feynman path integral for paths in $\R^n$. Here $G^\C$ is the group (under point-wise addition) of Gaussian\footnote{By Gaussian paths we mean the pointed paths are characterized by a mean and covariance.} pointed paths $X_a\ni (x,t_a)\rightarrow(\C^n,x_a)$ where $t_a\in\R$, $x(t_a)=x_a\in \R^n$. $X_a$ is an infinite-dimensional abelian topological group when endowed with a suitable topology. The corresponding path integral over $X_a$ is a formal object. But as soon as one imposes a constraint on the loose ends of the paths, for example $x(t_b)=x_b\in\R^n$ which `pins' them to a single point, the group `localizes' to a finite-dimensional Lie group $X_{a,b}\cong\R^n$.\footnote{To see this, parametrize the space of Gaussian pointed paths by mean and covariance. Fixing the loose end-point fixes the mean, and the covariance is then parametrized by points in $\R^n$. Consequently, the moduli space of Gaussian pointed paths with both end-points fixed in $\R^n$ is congruent to $\R^n$.} Being a finite-dimensional topological vector space, it is automatically locally compact: The corresponding path integral can now be explicitly evaluated.

 There are of course many other `constraints' that one can impose on a given system. These constitute the set $\Lambda$, and a particular choice of $\lambda\in\Lambda$ leads to a particular and explicit evaluation of the path integral over $X_a$.
\end{example}

Finally, to be economical, we suppose dynamics are modeled by \emph{inner} automorphisms of $\mathbf{F}(G^\C)$. It is natural to expect the algebra to include system dynamics since it contains a linear Lie group. After all, the Lie bracket induces a derivation on the algebra. This leads to our last assumption:
\begin{assumption}
The dynamics of a closed quantum system are governed by continuous {inner} automorphisms $Inn(\mathbf{F}(G^\C))$ generated by a dynamical group $G_D$ such that ${G_D}_\lambda\subseteq G_{\lambda}^\C$.
\end{assumption}

Expectations between physical states are ultimately determined by the dynamical group $G_D$ not $G^\C$.  Depending on the nature of $G_D$, the Hilbert space $\mathcal{H}$ determined by $G^\C$ may be restricted.  Later on we will investigate dynamics based on the Heisenberg and Poincar\'{e} groups. In both cases, $\mathcal{H}$ is restricted.

\subsection{Hilbert space of induced representations}\label{URs}
Suppose a family $G_{\Lambda}^\C$ that governs some quantum system has been identified.  The first order of business in the quantization program is to determine $\mathcal{H_{\lambda}}$ and find all relevant representations $\rho:G_{\lambda}^\C\rightarrow L_B(\mathcal{H_{\lambda}})$ for all $\lambda\in\Lambda$ where $L_B(\mathcal{H}_{\lambda})$ denotes the set of linear, bounded operators on $\mathcal{H}_{\lambda}$. Keeping in mind the important qualifiers indicated by sub/superscripts in $G^\C_{\lambda}$, we will simply write $G$ for notational clarity \emph{in this subsection} to indicate a complexified locally compact topological Lie group.

Consider the lie algebra $\mathfrak{G}$ of $G$. Our preliminary goal is to determine and interpret the representations $\varrho'^{(r)}:\mathfrak{G}\rightarrow L(\mathcal{V}^{(r)})$ furnished by $\mathfrak{G}$-modules $\mathcal{V}^{(r)}$ and labeled by $r$. Start with the triangular decomposition of the Lie algebra $\mathfrak{G}$;
\begin{equation}
\mathfrak{G}=\mathfrak{G}_-\oplus\mathfrak{G}_0\oplus\mathfrak{G}_+
\end{equation}
where
\begin{eqnarray}
&&\left[\mathfrak{G}_0,\mathfrak{G}_0\right]=0\notag\\
&&\left[\mathfrak{G}_+,\mathfrak{G}_-\right]\subseteq\mathfrak{G}_0\notag\\
&&\left[\mathfrak{G}_\pm,\mathfrak{G}_0\oplus\mathfrak{G}_\pm\right]
\subseteq\mathfrak{G}_\pm\;.
\end{eqnarray}

In dominant-integral lowest/highest-weight representations, the Cartan subalgebra $\mathfrak{G}_0$ defines \emph{potential} `quantum numbers' and `ground states' through a weight-space decomposition where `quantum numbers' correspond to weights in the basis of fundamental weights and `ground states' correspond to the lowest/highest weight vector.  These are \emph{potential} identifications, because we have not yet determined that the quantum framework will respect the Lie algebra structure.

Let $\varrho':\mathfrak{G}\rightarrow L(\mathcal{V})$ be a complex representation
with $\mathcal{V}$ a $\mathfrak{G}$-module. The
triangular decomposition of the algebra induces a decomposition of
$\mathcal{V}$ by
\begin{equation}
\mathcal{V}=\bigoplus_{w} \mathcal{V}_{(w)}
\,,\;\;\;\mathcal{V}_{(w)}:=\{\Bold{v}\in \mathcal{V}
:\varrho'(\mathfrak{h}_i)\Bold{v}=w_i\Bold{v}\}\,,\;\;\;i\in\{1,\ldots,\mathrm{rank}(G)\}
\end{equation}
where $\mathfrak{h}_i\in\mathfrak{G}_0$ and ${w}=\{w_1,\ldots,w_{\mathrm{rank}(G)}\}$
is a weight in the Dynkin basis composed of generally complex eigenvalues (since $\mathfrak{G}_0$ is complex in this case). Accordingly, representations are
partially characterized by $\mathrm{rank}(G)$ `quantum numbers'.

In the weight decomposition of \emph{finite-dimensional} $\mathcal{V}$, there is a distinguished subspace
$\mathcal{V}_{(w_+)}\subset\mathcal{V}$ associated with a maximal
weight ${w}_+$ such that
\begin{equation}\label{vacuum}
\varrho'(\mathfrak{g}_+)\Bold{v}=0,
\;\;\;\forall\mathfrak{g}_+\in\mathfrak{G}_+\,\mathrm{and}\,\forall\Bold{v}\in\mathcal{V}_{(w_+)}\;.
\end{equation}
The same can be said for minimal weights $w_-$.  For simplicity, we will assume the dynamical system is invariant under the inner automorphism $\mathfrak{G}_-\leftrightarrow\mathfrak{G}_+$, so there is no physical distinction between minimal and maximal weight. We might as well follow mathematics convention and confine our attention to maximal weight modules.\footnote{However, there may be interesting physics associated with dynamical systems that are not invariant under $\mathfrak{G}_-\leftrightarrow\mathfrak{G}_+$, and this case deserves investigation.}

In particular, if $\varrho'$ is an \emph{irreducible representation} (IR), then there is only one maximal weight (now called the highest weight) and
$\mathcal{V}_{(w_+)}$ is one-dimensional possessing a unique (up to scalar multiplication) highest-weight vector $\Bold{v}_{w_+}$. In this case module $\mathcal{V}$, now denoted $\mathcal{V}_{w_+}$, is called a highest-weight module, and it is generated by acting on
$\Bold{v}_{w_+}$ with combinations of `lowering operators'
$\mathfrak{g}_-\in\mathfrak{G}_-$. This of course is a familiar story.

Finite-dimensional, highest-weight IRs have three important properties: (i) if the highest weight is dominant-integral, then $\mathcal{V}_{w_+}$ possesses a positive-definite Hermitian inner product, (ii) the module furnishes an IR for the identity component of $G$ by exponentiation of $\varrho'$, and (iii) its highest-weight vector $\Bold{v}_{w_+}$ is a good candidate for a ground state.

Unfortunately highest-weight $\mathcal{V}_{w_+}$ will not generally be finite-dimensional for interesting algebras possessing a triangular decomposition. So these nice properties won't necessarily follow. Even worse, we will eventually want \emph{unitary} representations of (possibly) non-compact groups associated with $\mathfrak{G}$, and it is well-known that such representations are necessarily infinite dimensional. Fortunately there is a work-around --- induced representations.\footnote{The archetypal example of an induced representation is the little group method\cite{Wein} applied to the Poincar\'{e} group.}

Induced representations allow to leverage finite-dimensional irreducible representations of the maximal compact subalgebra $\mathfrak{G}_c\subseteq\mathfrak{G}$ to construct (possibly) infinite-dimensional, generically reducible representations of $G$. The relevant finite-dimensional IRs can be found first for the real form $\mathfrak{G}_c^\R$ of $\mathfrak{G}_c$ thanks to the fact that, given a finite-dimensional complex irreducible $\varrho'_\R:\mathfrak{G}^\R\rightarrow L(\mathcal{V})$, there exists a unique irreducible extension $\varrho':\mathfrak{G}\rightarrow L(\mathcal{V})$ such that $\varrho'(\mathfrak{g}_\C)=\varrho'_\R(\mathfrak{g}_1)+i\varrho'_\R(\mathfrak{g}_2)$ for all $\mathfrak{g}_\C=\mathfrak{g}_1+i\mathfrak{g}_2\,\in \mathfrak{G}$ with $\mathfrak{g}_1,\mathfrak{g}_2\in\mathfrak{G}^\R$.\cite[prop. 11.3]{B}

For finite or infinite dimensional $\mathcal{V}_{w_+}$, let $\mathcal{V}_{({\mu})}\subset\mathcal{V}_{w_+}$ denote the submodule
generated by $\mathfrak{G}^\R_c$ acting on a dominant-integral highest-weight vector
$\Bold{v}_{w_+}$. The submodule $\mathcal{V}_{({\mu})}$ furnishes a complex
IR of $\mathfrak{G}^\R_c$ and, by extension according to the cited proposition, it also furnishes a complex IR of $\mathfrak{G}_c$. Since $w_+$ is a highest weight,
$\mathcal{V}_{({\mu})}$ is an invariant subspace with respect to the parabolic subalgebra $\mathfrak{P}:=\mathfrak{G}_+\cup\mathfrak{G}_c$, i.e.
$\bar{\varrho}{\,'}(\mathfrak{P})\mathcal{V}_{({\mu})}\subseteq\mathcal{V}_{({\mu})}$ with $\bar{\varrho}{\,'}$ a sub-representation of $\varrho'$. This can be seen by using the triangular decomposition and the fact that $\mathfrak{G}_0\subseteq\mathfrak{G}_c$. Moreover, $\bar{\varrho}{\,'}$ exponentiates to an IR $\bar{\varrho}$ of the parabolic subgroup $P\subset G$ in the identity component since $\mathcal{V}_{({\mu})}$ is finite-dimensional.

So, for a given \emph{fundamental} (or \emph{basic})  irreducible representation $r$(see e.g. \cite[\S13.6]{FS}), we have a  highest-weight $\mathfrak{G}$-module $\mathcal{V}^{(r)}_{w_+}$  with an invariant submodule $\mathcal{V}^{(r)}_{(\mu)}$ under the parabolic subgroup $P\subset G$. Use this data to construct a principal coset bundle
$\mathcal{G}:=(G,X,\breve{pr},P)$ and its associated vector bundle
$(\mathcal{V}^{(r)}_{w_+},X,pr,\mathcal{V}^{(r)}_{(\mu)},P)$ where the base space
is the homogeneous coset space $X:=G/P$ and $\breve{pr}$ (respectively $pr$) denotes the principal (respectively vector) bundle projection. Finally, form the  Whitney-sum vector bundle for all relevant unitary (tensor product) representations furnished by the basic IR modules $\mathcal{V}^{(r)}_{(\mu)}$;
\begin{eqnarray}
\mathcal{W}_{\mathcal{V}}&:=&\bigoplus_{r=r_1}^{r_{max}}\big(\otimes\mathcal{V}^{(r)}_{w_+},X,pr,
\otimes\mathcal{V}^{(r)}_{(\mu)},P\big)\notag\\
&=:&\left(\mathcal{W},X,pr, \mathcal{W}_{(\Bold{\mu})},P\right)
\end{eqnarray}
where $\mathcal{W}_{(\Bold{\mu})}$ is a Hilbert module and $\Bold{\mu}:=(\mu^{(r_1)},\ldots,\mu^{(r_{max})})$.

Following standard methods, an induced unitary representation (UR) of $G$ is furnished by a vector space $\mathbf{E}_{P}^{G}$ of continuous, equivariant functions;
\begin{equation}\label{induction}
\mathbf{E}_{P}^{G}
:=\{\breve{\psi}\in
L^2({G},\mathcal{W}_{(\mu)})\,|\,\breve{\psi}(g\,p)=\bar{\varrho}(p^{-1})\breve{\psi}(g)\}
\end{equation}
where $p\in P$ and $\bar{\varrho}:P\rightarrow
L(\mathcal{W}_{(\mu)})$ is the IR obtained by exponentiation of $\bar{\varrho}'$. Note that $\mathbf{E}_{P}^{G}$ is a vector space under point-wise addition, and it can be given a Hilbert structure\cite{KT}. The induced representation $\mathrm{ind}_{P}^{G}\bar{\varrho}$ can be expressed as
\begin{equation}\label{representation}
(\mathrm{ind}_{P}^{G}\bar{\varrho}({g})\breve{\psi})(g_o)=\breve{\psi}(g^{-1}g_o)
=:\breve{\psi}_{{g}}(g_o)
\end{equation}
where $g_o,{g}\in {G}$. We stress \emph{these induced representations are not generally irreducible.}

It is trivial to check that $\mathrm{ind}_{P}^{G}\bar{\varrho}$ is a representation and that it is unitary with respect to
\begin{equation}\label{first inner product}
\langle\breve{\psi}_1|\breve{\psi}_2\rangle_{\mathbf{E}_{P}^{G}}
:=\int_G(\breve{\psi}_1(g)|\breve{\psi}_2(g))_{\mathcal{W}_{(\Bold{\mu})}}\;d\nu(g)\;.
\end{equation}
where $\nu$ is a \emph{left} Haar measure. Less trivially, \cite{KT} proves that $g\mapsto\mathrm{ind}_{P}^{G}\bar{\varrho}(g)\breve{\psi}$ is continuous, it respects the direct sum structure of $\mathcal{W}_{(\Bold{\mu})}$, and it is equivalent under group conjugation of $P$ (that is, $\mathrm{ind}_{g_oPg_o^{-1}}^{G}\bar{\varrho}(g_ogg_o^{-1})=\mathrm{ind}_{P}^{G}\bar{\varrho}(g))$). So $\mathrm{ind}_{P}^{G}\bar{\varrho}:G\rightarrow L_B(\mathbf{E}_{P}^{G})$ is a continuous, ad-invariant UR and $\mathbf{E}_{P}^{G}$ is Hilbert.

Crucially, when $\bar{\varrho}$ is unitary, $\langle\breve{\psi}_1|\breve{\psi}_2\rangle_{\mathbf{E}_{P}^{G}}$ is invariant under vertical automorphisms of $\mathcal{G}$. This is just the statement of gauge covariance (relative to $P$) in a fiber bundle framework. Insofar as a vertical automorphism can be viewed as a change of basis in $\mathcal{W}_{(\mu)}$ and the choice of a fiducial basis is arbitrary, we insist that gauge covariant $\breve{\psi}\in\mathbf{E}_{P}^{G}$ correspond to kinematical quantum states.

It is customary and useful to express $\mathbf{E}_{P}^{G}$ in terms of a space of suitable sections of the vector bundle $\mathcal{W}_{\mathcal{V}}$. Start with \emph{normalized} equivariant functions $\breve{\psi}\in
L^2({G},\mathcal{W}_{(\mu)})$
\begin{equation}\label{normalized induction}
\mathbf{NE}_{P}^{G}
:=\{\breve{\psi}\in
L^2({G},\mathcal{W}_{(\mu)})\,|\,\breve{\psi}(g\,p)=N(p)\bar{\varrho}(p^{-1})\breve{\psi}(g)\}
\end{equation}
where the normalization is given by $N^2(p):={\triangle_{P}(p)/
\triangle_{G}(p)}$ with
$\triangle_G(g)=|\mathrm{det}\,Ad_G(g)|$ the modular function of
$G$, and $\bar{\varrho}:P\rightarrow
L(\mathcal{W}_{(\mu)})$ is a \emph{unitary} sub-representation of $\varrho$.
The induced representation $\mathrm{ind}_{P}^{G}\bar{\varrho}:G\rightarrow L_B(\mathbf{NE}_{P}^{G})$ is again expressed as
\begin{equation}\label{normalized representation}
(\mathrm{ind}_{P}^{G}\bar{\varrho}({g})\breve{\psi})(g_o)=\breve{\psi}(g^{-1}g_o)
=:\breve{\psi}_{{g}}(g_o)
\end{equation}
where $g_o,{g}\in {G}$.

Recall that $\breve{\psi}\in L^2({G},\mathcal{W}_{(\mu)})$ and sections $\psi\in L^2(X,\mathcal{W})$ can be identified\footnote{Because the principal and vector bundles are associated bundles, $g$ is both a point in $\breve{pr}^{-1}(x)\in G$ and an admissible map $g:\mathcal{W}_{(\Bold{\mu})}\rightarrow pr^{-1}(x)\in\mathcal{W}$ \cite[ pg. 367]{C-B}.} according to  $\breve{\psi}(g)=g^{-1}\circ\psi(x)$ with the condition that $\breve{pr}(g)=x=gx_0$
where $x_0$ is a choice of origin in $X$. If a canonical local
section $s_i$ on the principal bundle is chosen relative to a
local trivialization $\{U_i,\varphi_i\}$, then $\breve{\psi}$ and
$\psi$ are in fact \emph{canonically} related, and we can identify $
\psi(\cdot)\equiv\breve{\psi}(s_i(\cdot))$: In other words $\psi\equiv s_i^\ast\breve{\psi}$. Explicitly, for $x\in U_i\subset X$, the
representative of $\psi(x)$ relative to the local trivialization is
${\psi}(x)=(x,\Bold{v}_{{w}_g})$ where
$\Bold{v}_{{w}_g}=\breve{\psi}(g)$. This canonical identification is
tantamount to a particular choice of basis for each fiber
$\mathcal{W}_x\subset\mathcal{W}$ over $U_i$ given by
$\Bold{e}_x=(x,\Bold{e}_{({\mu})})$ where
$\Bold{e}_{({\mu})}$ is a basis of $\mathcal{W}_{(\Bold{\mu})}$.

Note that $\breve{pr}(gs_i(x))=g\,\breve{pr}(s_i(x))=gx$ so $gs_i(x)$
must be a point in the fiber over $gx$, i.e.
$gs_i(x)=s_i(gx)\tilde{p}$ for some $\tilde{p}\in P$.  From its defining relation, $\tilde{p}$ depends on both $x$ and $g$ which we will emphasize by writing $\tilde{p}(x,g)$. It satisfies
\begin{eqnarray}\label{cocycle}
\tilde{p}(x_0,p)&=&s_i(x_0)^{-1}\,p\,s_i(x_0)\;,\;\;\;\forall p\in P\notag\\
\tilde{p}(x,e)&=&e\;,\;\;\;\forall x\in G/P\notag\\
\tilde{p}(x,g_1g_2)&=&\tilde{p}(g_2 x,g_1)\tilde{p}(x,g_2)\;.
\end{eqnarray}
In particular, the last two imply $\tilde{p}(x,g^{-1})^{-1}=\tilde{p}(g^{-1}x,g)$ taking $g_1=g$ and $g_2=g^{-1}$.

Hence, using canonical local sections, we can define a
canonical induced representation $\rho:G\rightarrow L(L^2(X,\mathcal{W}))$ by $\rho(g)\circ s_i^\ast=s_i^\ast\circ\mathrm{ind}_{P}^{G}\bar{\varrho}(g)$. Explicitly,
\begin{eqnarray}\label{rho representation}
(\rho({g}){\psi})(x)&:= &(\mathrm{ind}_{P}^{G}\bar{\varrho}({g})\breve{\psi})(s_i(x))\notag\\
&=&\breve{\psi}(g^{-1}s_i(x))\notag\\
&=&\breve{\psi}(s_i(g^{-1}x)\tilde{p}(x,g^{-1}))\notag\\
&=&N(\tilde{p}(x,g^{-1})){\bar{\varrho}}(\tilde{p}(x,g^{-1})^{-1})s_i^\ast\breve{\psi}(g^{-1}x)\notag\\
&=&N(\tilde{p}(g^{-1}x,g)^{-1}){\bar{\varrho}}(\tilde{p}(x,g^{-1})^{-1})\psi(g^{-1}x)\notag\\
&=&N(\tilde{p}(g^{-1}x,g))^{-1}{\bar{\varrho}}(\tilde{p}(g^{-1}x,g))\psi(g^{-1}x)
\end{eqnarray}
where the last line follows since $N$ is a homomorphism $N:P\rightarrow\R_+$.

Evidently the action of $\rho(G)$ on $\psi$ is determined  only modulo the right action of $\tilde{p}(x,g)$. This ambiguity comes from the ambiguity in the basis (relative to $\mathcal{W}_{(\Bold{\mu})}$) for each fiber $\mathcal{W}_x$ inherited by a choice of local trivialization. Accordingly, physical $\psi$ should be covariant under vertical automorphisms $V:\mathcal{G}\rightarrow \mathcal{G}$ by $V(g)=g\upsilon(g)$ where $\upsilon:G\rightarrow P$. This is consistent with $\psi\equiv s_i^\ast\breve{\psi}$ and the earlier observation that $\breve{\psi}$ likewise should be covariant.

Hence, we posit that gauge-covariant quantum states can be represented by an equivalence class $[\psi]$ where the equivalence relation is
\begin{equation}
\psi(x)\sim V^\ast\psi(x)=N(\upsilon(s_i(x)))\bar\varrho(\upsilon(s_i(x))^{-1})\psi(x)\;.
\end{equation}
Explicitly, in a local trivialization this is
\begin{equation}\label{equivalence relation}
(x,\Bold{v}_{{w}_{g}})\sim(x,N(p_{s_i(x)})\bar{\varrho}(p_{s_i(x)}^{-1})\Bold{v}_{{w}_g})
\end{equation}
where $p_{s_i(x)}:=\upsilon(s_i(x))\in P$ and $\Bold{v}_{{w}_g}=\breve{\psi}(g)$. So, relative to a canonical section, gauge covariance can be expressed as $(x,\breve{\psi}(g))\sim (x,\breve{\psi}(gp_{s_i(x)}))$ for all $x\in X$. This is just the coset space version of the covariance condition on $\breve{\psi}\in\mathbf{E}^G_P$ .

Use the Hermitian inner product $(\cdot|\cdot)$ on
$\mathcal{W}_{(\Bold{\mu})}$ and the left Haar measure $\nu$ on $G$ to construct a bundle metric on $\mathcal{W}$.
Then equip $L^2(X,\mathcal{W})$ with a Hermitian inner product induced from
$\mathcal{W}$ and the quasi-invariant measure on $X$ given by $d\nu_{P}(g^{-1}x)=N^2(\tilde{p}(x,g^{-1}))d\nu_{P}(x)$ according to
\begin{equation}\label{inner product}
\langle\psi_1|\psi_2\rangle=\int_X(\psi_1(x)|\psi_2(x))_{\mathcal{W}_x}\;d\nu_{P}(x)\;.
\end{equation}
It is easy to see that $\langle\rho(g)\psi_1|\rho(g)\psi_2\rangle=\langle\psi_1|\psi_2\rangle$, and it is invariant under vertical automorphisms of $\mathcal{G}$. Hence $\rho$, like $\mathrm{ind}_{P}^{G}\bar{\varrho}$, is a continuous, ad-invariant representation that is unitary with respect to the given inner product. Complete $L^2(X,\mathcal{W})$ with respect to the associated norm. Then
$\mathcal{H}\equiv L^2(X,\mathcal{W})$ is Hilbert and it models the kinematical quantum Hilbert space by postulate.

It must be emphasized that $\mathcal{H}$ does not necessarily coincide with the \emph{physical} Hilbert space $\mathcal{H}_D$ generated by the assumed dynamical group $G_D\subseteq G$.
\begin{definition}\label{physical Hilbert space}
The physical Hilbert space is $\mathcal{H}_D=L^2(X_D,\mathcal{W}_D)$ endowed with the inner product \emph{(\ref{inner product})} where now ${\psi}\in L^2(X_D,\mathcal{W}_{D})$ and ${\mathcal{W}_{D}}_{(\mu)}$ furnishes (a possible tensor product of) a direct sum of all relevant IRs of $P_D\subset G$.
\end{definition}
If the dynamical group ${G_D}$ is connected and simply connected with $H^2({G_D},\R)\cong0$, then there are no intrinsic projective representations. Otherwise, one must find a projective representation, or include central charges, or simply promote $G_D$ to its universal cover.

To ensure gauge equivalence of the eventual quantum system with respect to the dynamical group, it is necessary that observables associated with each $p\in P_D$ act on $\mathcal{H}_D$ as symmetries with conserved quantities. By construction, vectors in ${\mathcal{W}_{D}}_{(\mu)}$, which correspond to physical, local quantum states $\psi(x)$, are labeled by their gauge equivalence classes and weights/quantum numbers associated with IRs of $G_{c}$; and  the physical ground state $\psi_0$ can be defined by its representative $\psi_0(x)=(x,\Bold{v}_{w_+})$ for all $x\in X$ (equivalently, $\breve{\psi}_0(g)=\Bold{v}_{w_+}$ for all $g\in G$). As the inner product is invariant under $\mathrm{ind}_{P}^{G}\bar{\varrho}(P_D)$ and vertical automorphisms of $\mathcal{G}$, it is legitimate to identify  physical-state weights/quantum numbers with conserved charges up to (re)normalization as long as $P_D$ is contained in the $C^\ast$-algebra and it commutes with the generator of system evolution.

Now that we possess $G_D\subseteq G$, its restricted induced URs $\rho_D$ coming from $\rho$, and the physical Hilbert space $\mathcal{H}_D=L^2(X_D,\mathcal{W}_D)$; it remains to construct a suitable $C^\ast$-algebra of functionals to model $\mathfrak{A}_L$.

\subsection{Functional Mellin transform}
To construct the $C^\ast$-algebra we use the functional Mellin transform which is a particular type of functional integral defined in \cite{LA3}. Roughly, to define functional integrals in general, we take the data $(G,\mathfrak{B},G_{\Lambda})$  with $\mathfrak{B}$ an associative  Banach algebra  and define a functional integral by a \emph{family} of integral operators $\mathrm{int}_\Lambda:\mathbf{F}(G)\rightarrow \mathfrak{B}$ where $\mathbf{F}(G)$ is a certain  space of continuous morphisms\footnote{$M(\mathfrak{B})$ denotes the multiplier algebra of $\mathfrak{B}$. Normally, for physics applications $\mathfrak{B}$ will be a unital algebra in which case $M(\mathfrak{B})=\mathfrak{B}$.} $\mathrm{F}\in \mathrm{Mor}_C(G,M(\mathfrak{B}))$ whose restrictions $f:=\mathrm{F}|_{G_\lambda}$ are Haar-integrable for all $\lambda\in\Lambda$. A brief introduction is given in appendix \ref{appx. A}.

In particular, functional Mellin transforms are defined using the refined functional integral data $(G^\C,\mathfrak{C}^\ast,G^\C_{\Lambda})$ where $\mathfrak{C}^\ast$ is a \emph{unital} $C^\ast$-algebra and $G^\C$ is a \emph{path connected} complex topological group.
 \begin{definition}\emph{(\cite{LA2})}\label{Mellin def.}
Let $\rho:G^\C_{\lambda}\rightarrow
\mathfrak{C}^\ast$  be a strictly-continuous, injective homomorphism, and let
$\pi:\mathfrak{C}^\ast\rightarrow L_B(\mathcal{H})$ be a representation. Define a subspace of integrable
functionals ${\widetilde{\mathbf{F}}}(G^\C)\subseteq{\mathbf{F}}(G^\C)$ such that $\mathrm{F}\in \mathrm{Mor}_C(G^\C,\mathfrak{C}^\ast)$ is
equivariant under right-translations according to $\mathrm{F}(gh)=\mathrm{F}(g)\rho(h)$. The functional Mellin
transform
$\mathcal{M}_\lambda:\widetilde{\mathbf{F}}(G^\C) \rightarrow \mathfrak{C}^\ast$ is
defined by
\begin{equation}\label{Functional Mellin}
\mathcal{M}_\lambda\left[\mathrm{F};\alpha\right]
:=\int_{G^\C}\mathrm{F}(gg^{\alpha})\;\mathcal{D}_\lambda g=\int_{G^\C}\mathrm{F}(g)\rho(g^{\alpha})\;\mathcal{D}_\lambda g
\end{equation}
where $\alpha\in\mathbb{S}\subset\C$, $g^\alpha:=\exp_G(\alpha\log_Gg)$, and $\pi(F(g)\rho(g^\alpha))\in L_B(\mathcal{H})$ where the space of bounded linear operators $L_B(\mathcal{H})$ is given the strict topology.
Denote the space of Mellin integrable functionals by $\mathbf{F}_{\mathbb{S}}(G^\C)$.
\end{definition}

The right-hand side of (\ref{Functional Mellin}) is given meaning through the definition of functional integral (see appx. \ref{appx. A}), so we need to define the Mellin transform for a locally compact topological group:
\begin{definition}
Let the map $\rho:G_{\lambda}^\C\rightarrow \mathfrak{C}^\ast$  be a strictly-continuous injective
homomorphism, and consider equivariant functions  $g^{1+\alpha}\mapsto f(g)\rho(g^\alpha)$  such that $f=\mathrm{F}|_{G^\C_\lambda}\in C_c(G_{\lambda}^\C,\mathfrak{C}^\ast)$  for all $\alpha\in
\mathbb{S}\subset\C$. Then $\mathrm{F}$ is Mellin integrable if
\begin{equation}\label{Mellin}
  \left|\mathcal{M}_\lambda\left[\mathrm{F};\alpha\right]\right|\leq\int_{G^\C_{\lambda}}|f(g_\lambda)\rho(g_\lambda^\alpha)|
  \,d\nu(g_\lambda)<\infty\,,\;\;\;\;\alpha\in\mathbb{S}\;.
\end{equation}
We say the Mellin transform $\mathcal{M}_\lambda\left[\mathrm{F};\alpha\right]$ exists in the fundamental region $\mathbb{S}$.

Identifying the Lie algebra $\mathfrak{G}^\C$ of $G^\C$ at the identity element with the space of complex one-parameter subgroups $\mathfrak{L}(G^\C)$, the Mellin integral can be explicitly formulated as
\begin{eqnarray}\label{mellin integral}
  \int_{G_{\lambda}^\C}f(g_\lambda)\rho(g_\lambda^\alpha)
  \,d\nu(g_\lambda)\notag\\
  &&\hspace{-1.75in}=\int_{\mathfrak{L}(G_{\lambda}^\C)}f(\exp_{G_{\lambda}^\C}(\mathfrak{g}))
  \rho(\exp_{G_{\lambda}^\C}(\alpha\mathfrak{g}))\,
  |\det\,d_{\mathfrak{g}}\exp_{G_{\lambda}^\C}(\mathfrak{g})|\;d\mathfrak{g}\;.
\end{eqnarray}

\end{definition}
Succinctly, the functional Mellin transform is a family of integrals represented by the right-hand side of (\ref{mellin integral}) which can be interpreted as a generalized two-sided Laplace transform.

\begin{remark}
The class of functional Mellin transforms defined here includes,  as a special case, the integrated form of a covariant representation $(\pi,U)$ of a dynamical system that characterizes crossed products\emph{\cite{W}}. The integrated form of $(\pi,U)$, denoted $\pi\rtimes
U(f)$, is essentially equivalent to $\pi(\mathcal{M}_\lambda\left[\mathrm{F} ;1\right])$ provided $\pi\circ\rho$ is a strongly-continuous unitary
representation $\pi\circ\rho:G^\C_{\lambda}\rightarrow L_B(\mathcal{H})$.
\end{remark}

One can show that $(\mathrm{F}_1\ast\mathrm{F}_2)(gh)=(\mathrm{F}_1\ast\mathrm{F}_2)(g)\rho(h)$ using the $\ast$-convolution product in $\mathbf{F}(G^\C)$. Hence $\widetilde{\mathbf{F}}(G^\C)$ is a subalgebra, and one defines a norm on the associated space of Mellin integrable functionals $\mathbf{F}_{\mathbb{S}}(G^\C)$ by $\|\mathrm{F}\|_{\mathbb{S}}
:=\mathrm{sup}_{\alpha,\lambda}\|\mathcal{M}_\lambda\left[\mathrm{F};\alpha\right]\|$ with $\alpha\in\mathbb{S}$
and  then completes $\mathbf{F}_{\mathbb{S}}(G^\C)$  w.r.t. this (or some other suitably defined) norm. Then,
\begin{proposition}\emph{(\cite{LA2} Th. 3.20)}
For suitably restricted $\alpha\in\mathbb{S}$, the space $\mathbf{F}_{\mathbb{S}}(G^\C)$ is a $C^\ast$-algebra when
 endowed with an involution defined by $\mathrm{F}^\ast(g^{1+\alpha}):=\mathrm{F}(g^{-1-\alpha})^\ast\Delta(g^{-1})$
 and equipped with a suitable topology.
\end{proposition}

The utility of functional Mellin transforms is they realize $\ast$-representations of $\mathbf{F}_{\mathbb{S}}(G^\C)$ under suitable conditions.\cite{LA2} For example, if $\mathfrak{C}^\ast$ is commutative then $\mathcal{M}_\lambda$ is a $\ast$-representation for all $\alpha\in\mathbb{S}$. On the other hand, if $\mathfrak{C}^\ast$ is non-commutative but $G^\C$ is abelian, then $\mathcal{M}_\lambda$ is a $\ast$-representation for $\alpha\in\R\cap\mathbb{S}$ if $\rho$ is unitary or $\alpha\in i\R\cap\mathbb{S}$ if $\rho$ is real. In the extreme case of non-commutative $\mathfrak{C}^\ast$ and non-abelian $G^\C$, then $\mathcal{M}_\lambda$ is a $\ast$-representation only for $\alpha\in\{0,1\}$ with $\rho$ unitary (or $\alpha\in\{- i,i\}$ with $\rho$ real). We denote these separate cases by a single symbol $\mathit{\Pi}_\lambda^{(\alpha)}$ where $\alpha$ must be determined by context.

The upshot is $\mathbf{F}_{\mathbb{S}}(G^\C)$ and $\mathcal{M}_\lambda$ offer a practical means to effect quantization given a strictly-continuous representation $\rho:G_{\lambda}^\C\rightarrow \mathfrak{C}^\ast$. In the context of QM, we will: 1) insist that $\rho$ is a UR induced from $P$ according to the previous subsection; 2) identify $\mathfrak{C}^\ast\equiv L_B(\mathcal{H})$; and 3) equip $L_B(\mathcal{H})$ with the strong topology.  With this understood, the kinematical framework is nearly complete --- it remains to interpret the set of homomorphisms $\Lambda$.

\subsection{Observation/Measurement}
Although the act of observation/measurement is sometimes interpreted as non-unitary evolution --- and hence dynamical in nature --- in this scheme it is more naturally interpreted as kinematical.

We have seen that the functional $\mathrm{F}\in \mathbf{F}_{\mathbb{S}}(G^\C)$ loosely corresponds to an entire family of functions $f\in L^1(G_{\lambda}^\C,L_B(\mathcal{H}))$ for each $\lambda\in\Lambda$ representing a set of homomorphisms onto measurable topological groups. It is easy to imagine that quantum states as well as the quantum states of macroscopic observers that actualize some observable\footnote{As usual, an observable is a self-adjoint element $\mathrm{O}\in\mathbf{F}_{\mathbb{S}}(G^\C)$.} are modeled by suitable families of these functions.

Furthermore, the convolution products in $\mathbf{F}_{\mathbb{S}}(G^\C)$ and $L^1(G_{\lambda}^\C,L_B(\mathcal{H}))$ are equivalent by definition, but their respective norms are not. Our choice of norm on $\mathbf{F}_{\mathbb{S}}(G^\C)$ --- along with the fact that the convolution product and involution are only defined within each $L^1(G_{\lambda}^\C,L_B(\mathcal{H}))$ --- renders its restriction a direct sum $\mathbf{F}_{\mathbb{S}}(G^\C)|_{G_{\Lambda}}=\bigoplus_{\lambda\in\Lambda} L^1(G_{\lambda}^\C,L_B(\mathcal{H}))$ if $G_\Lambda$ is a finite family \cite[pg. 16--17]{BEH}. Therefore, since we stipulate $G_\Lambda$ is a finite family, a `query' (which picks out a single $\lambda$) induces a projection.

So the measurement process gets a topological interpretation: Performing an observation and thereby actualizing an observable
corresponds to a particular projection from $\mathbf{F}_{\mathbb{S}}(G^\C)|_{G_{\Lambda}}$ to
$L^1(G_{\lambda}^\C,L_B(\mathcal{H}))$. Precisely which projection is effected cannot be known \emph{a priori}. Further, by assumption evolution of a closed system is generated by $G_D$, so any subsequent observation will be referred to $L^1(G_{D_{\lambda}},L_B(\mathcal{H}))\subset\mathbf{F}_{\mathbb{S}}(G^\C)$ unless \emph{external}\footnote{Since a closed system is supposed to evolve according to a known $G_{D_{\lambda}}\subseteq G_\lambda^\C$, it takes something outside the system to induce a new localization $\widetilde{\lambda}:G^\C\rightarrow G_{\widetilde{\lambda}}^\C$.} interaction takes the system out of this subspace.

\begin{example}
Return to the Feynman path integral example. Recall that observing a point-to-point transition leads to a `localization' $\lambda:X_a\rightarrow X_{a,b}$ to some finite-dimensional group $X_{a,b}$ --- typically $X_{a,b}\cong\R^3$. But there is no guarantee the identity in $X_a$ maps to the identity in $X_{a,b}$.\footnote{This is reflected in the functional integral by the left-invariance of the Haar measure.} In other words, in the typical case, there is no preferred origin in $X_{a,b}\cong\R^3$ until an observation/measurement selects one through a specific choice of $\lambda\in\Lambda$. Once selected, the group $X_{a,b}$ with its preferred origin continues to govern the \textbf{closed} system evolution. However, external interaction necessarily implies a new $G_{\tilde{\lambda}}^\C$ along with the concomitant localization ambiguity of the origin.
\end{example}

Essentially,  the topological aspect of the model supplies a measurable family of Hilbert spaces. The family represents indeterminacy; not of the system but of the observer. Once a measurement has been made, it is given comparative meaning through a specific representation of the associated observable $\mathcal{M}_\lambda[\mathrm{O};\alpha]$ furnished by
   the Hilbert space based on the locally compact group $G_{\lambda}$  that was selected by the observer. The framework contains and separates both ontic and epistemic notions, and it offers a topological replacement for wave function collapse.

\subsection{Quantum Hilbert module}
The previous subsections can be efficiently organized and expressed under the rubric of Hilbert $C^\ast$-modules (see e.g. \cite{LANDS1}). Suppose we know the topological Lie group $G^\C$ and algebra $\mathfrak{A}_L$ of some quantum system. Construct the vector bundle $(\mathcal{W},X,pr,\mathcal{W}_{(\mu)},P^\C)$ and its associated principal bundle $(G^\C,X,\breve{pr},P^\C)$ where $X=G^\C/P^\C$, the typical fiber $\mathcal{W}_{(\mu)}$ is Hilbert and possibly infinite-dimensional, and $P$ is some chosen parabolic subgroup.

Call $\mathcal{E}\rightleftharpoons\mathfrak{Q}^\ast$ a pre-quantum Hilbert $C^\ast$-module. It is comprised of a linear\footnote{The homomorphisms are required to map the identity in $G^\C$ to the origin in $\mathcal{W}_{(\mu)}$.} space $\mathcal{E}=\mathrm{Hom}_C(G^\C,\mathcal{W}_{(\mu)})$ and an algebra $\mathfrak{Q}^\ast=\mathbf{F}_{\mathbb{S}}(G^\C)$. Here $\mathbf{F}_{\mathbb{S}}(G^\C)$ is understood to be defined in terms of the functional Mellin transform. Identifying $\mathcal{E}$ with $L^2(X,\mathcal{W})$, the Hilbert module $L^2(X,\mathcal{W})\rightleftharpoons\mathbf{F}_{\mathbb{S}}(G^\C)$ underlies the kinematic backdrop of the quantum system.

But quantization requires one more step: Transition amplitudes between states in $\mathcal{E}$ only acquire meaning through a `localization' $\lambda:G^\C\rightarrow G^\C_\lambda$ where $G^\C_\lambda$ is locally compact. Then $\mathcal{E}$ and $\mathfrak{Q}^\ast$ acquire explicit representations $\mathcal{E}_\lambda=L_\lambda^2(X,\mathcal{W})$ and $\mathfrak{Q}^\ast_\lambda=\mathbf{F}_{\mathbb{S}}(G^\C_\lambda)$ through {induced} representations of $G^\C_\lambda$. So the {quantization} of a system characterized by $\mathcal{E}\rightleftharpoons\mathfrak{Q}^\ast$ is modeled by a \emph{family} of Hilbert $C^\ast$-modules $\mathcal{E}_\Lambda\rightleftharpoons \mathbf{F}_{\mathbb{S}}(G^\C_\Lambda)$. This framework allows one to work at an abstract level as opposed to the concrete level of the previous sections. Of course, it is important to have both levels at one's disposal.

\section{System evolution}
Recall that $\rho$ is a UR of $G_{\lambda}^\C$, and it is strictly not defined on $G^\C$. However, it is cumbersome and messy to keep indicating the $\lambda$ dependence by always writing $\rho(g_\lambda)$. So for this entire section we will simply write $\rho(g)$, but we will continue to distinguish between $G^\C$ and $G_\lambda^\C$.

\subsection{Hamiltonians}\label{Hamiltonians}
One lesson learned in \cite[ex. 4.2]{LA2} is that functional inverse powers $A^{-\alpha}_\lambda$ associated with one-parameter subgroups are characterized (under localization) by logarithmic-type maps on $\C^\times$. The logarithmic-type maps are rooted in the functional Mellin presentation of $A^{-\alpha}_\lambda$, and the localization is a natural choice since invertible $A$ excludes vanishing eigenvalues. Further, for \emph{self adjoint} functional inverse powers, the natural localization reduces to the real subgroup $\R_+\times\{1-,1\}$. In the case at hand, such an operator gives rise to a propagator associated with an evolution operator, and the logarithmic-type maps localize not to a time parameter that tracks ticks on an observer's clock but to an evolution parameter that tracks the interval between ticks on an observer's clock. To emphasize; the evolution-time parameter in our view represents an interval characterized by the duration of a dynamical process relative to some observer's clock.

To model quantum evolution, let $\phi_{\,\mathfrak{h}_{\mathrm{U}}}\in\mathrm{Hom}_C(\R,U(\mathbf{F}_{\mathbb{S}}(G^\C)))$ denote a continuous one-parameter subgroup of unitary units generated by a \emph{time-independent} observable $\mathfrak{h}_{\mathrm{U}}\in\mathfrak{G}^\C$. For $g\in\phi_{\,\mathfrak{h}_{\mathrm{U}}}(\R)$, define the functional $\mathrm{E}^{-\mathrm{H}}(g):=e^{-\mathrm{H}(g)}\in L_B(\mathcal{H})$. It will be convenient to break with convention by taking $\mathrm{H}(g)$ skew-Hermitian. Adopt the topological localization $\lambda_{\mathfrak{h}_{\mathrm{U}}}:\phi_{\,\mathfrak{h}_{\mathrm{U}}}(\R)\rightarrow \R_+\times\{i,-i\}\equiv i\R_+\cup i\R_-$ and choose functional $\mathrm{H}(g)=H\rho(g)=H\tau$ with \emph{imaginary} $\tau\in i\R_+\cup i\R_-$ and Hermitian $H\in L(\mathcal{H})$.  Then $\rho(g^\alpha)=e^{\alpha\rho'(\log(g))}=\tau^\alpha$ with the principle branch for $\log$. But $\rho'(\log(g)$ is skew-Hermitian\footnote{To see this note that $\log(\tau)\in i\R$ implies $(\log(\tau))^\dag=-\log(\tau)=\log(\tau^{-1})$. Hence,  for $\tau=(|\tau|,i)\in i\R_+$, conjugate-evolution-time $\tau^\dag=\tau^{-1}=(|\tau|^{-1},-i)\in i\R_-$ is \emph{not} the complex conjugate of $\tau$.} so we must have $\alpha\in\R\cap\mathbb{S}$ to ensure unitary $\rho$. Happily, this is a sufficient conditions\cite[lemma 3.16]{LA2} for functional Mellin to be a $\ast$-representation on $\mathcal{H}$ since $\phi_{\,\mathfrak{h}_{\mathrm{U}}}$ is abelian. Evidently, $\mathrm{H}(g)$ generates evolution  for $\tau\gtrless i0$ and $\tau^\dag\lessgtr i0$ where $\tau$ represents evolution-time \emph{intervals}. We interpret unitary $e^{-\mathrm{H}(g)}\in L_B(\mathcal{H})$ for all $\tau\in i\R_+\cup i\R_-$ as the union of two semi-groups that generate closed system dynamics in evolution-time and conjugate-evolution-time, but note that conjugate-evolution-time does \emph{not} match the usual notion of time reversal.

Accordingly, quantum evolution requires continuous, unitary inner automorphisms on $\mathbf{F}_{\mathbb{S}}(G^\C)$ of the form $\mathrm{E}^{-\mathrm{H}}$ where $\mathrm{E}$ is defined in terms of the $\ast$-convolution that represents multiplication, and $\mathrm{H}:=(\mathrm{Log}\,\mathrm{E}^{-\mathrm{H}})^{-1}$ is skew-adjoint where the functional $\mathrm{Log}\,\mathrm{F}\in\mathbf{F}_{\mathbb{S}}(G^\C)$ is defined by\cite{LA2}
\begin{equation}
(\mathrm{Log}\,\mathrm{F})^{-1}_\lambda
:=\left.\frac{d}{d\alpha}\mathcal{M}_\lambda
\left[\mathrm{E}^{-\mathrm{F}};\alpha\right]\right|_{\alpha\rightarrow0^+}
\end{equation}
if the limit exists.

Given an observable $\mathrm{H}$, functional Mellin can be used to construct various related observables in $\mathbf{F}_{\mathbb{S}}(G^\C)$ (under conditions that render the Mellin transform a $\ast$-representation). For example, $\mathit{\Pi}^{(\alpha)}_{\lambda}(\mathrm{E}^{-(\mathrm{H}-z\mathrm{Id})})$ localized by $\lambda_{\mathfrak{h}_{\mathrm{U}}}$ represents the resolvent of $\mathrm{H}$ if the evolution generator $\mathrm{H}(g)=H\tau$ is linear in $\tau$ where $H\in L(\mathcal{H})$, $\alpha\in\R\cap\mathbb{S}$, and $z\notin\sigma(H)$. This readily follows from functional Mellin\cite{LA2}. For an eigenbasis $|i\rangle\in\mathcal{H}$ of $H$,
\begin{eqnarray}\label{resolvent example}
\langle i|\mathit{\Pi}^{(\alpha)}_\lambda(\mathrm{E}^{-(\mathrm{H}-z\mathrm{Id})})|j\rangle
&=& \int_{\phi_{\,\mathfrak{h}_{\mathrm{U}}}(\R)} \langle i|e^{-
(\mathrm{H}-z\mathrm{Id})(g)}|j\rangle\,\rho(g^\alpha)\;\mathcal{D}_\lambda g\notag\\
&&\hspace{-1.5in}\stackrel{\lambda_{\mathfrak{h}_{\mathrm{U}}}}
{\longrightarrow}\frac{1}{\Gamma(\alpha)}\int_{0}^{i\infty}\langle i|e^{-
(H-zId)\tau}|j\rangle\,\tau^\alpha\;d\log (\tau)
+\frac{1}{\Gamma(\alpha)}\int_{0}^{-i\infty} \langle i|e^{-(H-zId) \tau}|j\rangle\,\tau^\alpha\;d\log (\tau)\notag\\
&&\hspace{-1.5in}=\frac{1}{\Gamma(\alpha)}\left(\int_{0}^{i\infty}+\int_{0}^{-i\infty}\right) \langle i|e^{-
(H-zId)\tau}|j\rangle\,\tau^\alpha\;d\log (\tau)
=\langle i|(H-zId)_{\Gamma}^{-\alpha}|j\rangle\;,
\;\;\alpha\in\langle 0,\infty\rangle\;.\notag\\
\end{eqnarray}
The second line uses the measure $\nu(g_\Gamma)=\log (g)/\Gamma(\alpha)$, and the third line uses left-invariance of the Haar measure and the fact that $\langle i|(H-zId)|j\rangle$ is invertible for $z\notin\sigma(H)$. Note that only one of the integrals will contribute depending on whether $\Im(\sigma(H-zId))\gtrless0$.

Similarly, in the same context, one can represent the functional inverse power\cite{LA2}
\begin{equation}\label{functional power}
H^{-\alpha}_\lambda:=\mathit{\Pi}^{(\alpha)}_\lambda(\mathrm{E}^{-\mathrm{H}})
=\int_{{\phi_{\,\mathfrak{h}_{\mathrm{U}}}(\R)}} e^{-Hg}\,g^\alpha\;\mathcal{D}_\lambda g\,,\;\;\;\;\alpha\in\mathbb{S}\;.
\end{equation}
Armed with the functional inverse power of some $\mathrm{H}\in\mathbf{F}_{\mathbb{S}}(G^\C)$, under suitable conditions on the spectrum of the associated operator $H\in L(\mathcal{H})$  one can define the functional trace and functional determinant (with suitable regularization)\cite{LA2};
\begin{equation}
\mathrm{Tr}\,{H}^{-\alpha}_\Gamma:=\frac{1}{\Gamma(\alpha)}\int_{i\R_+\cup i\R_-} \mathrm{tr}\left(e^{-H\tau}\,\tau^\alpha\right)\;d\log (\tau)\,,\;\;\;\;\alpha\in\mathbb{S}
\end{equation}
where $\mathrm{tr}$ is the trace in $\mathcal{H}$, and
\begin{eqnarray}
\mathrm{Det}\,{H}^{-\alpha}_\Gamma
:=\frac{1}{\Gamma(\alpha)}\int_{i\R_+\cup i\R_-} \mathrm{det}\left(e^{-H\tau}\,\tau^\alpha\right)\;d\log (\tau)
\,,\;\;\;\;\alpha\in\mathbb{S}\;.\notag\\
\end{eqnarray}

\begin{example}\label{Klein-Gordon}
With an eye toward Poincar\'{e} QM presented in section \emph{\ref{Poincare}}, let us represent the functional inverse power of the Klein-Gordon operator\footnote{The operator acts on $L_B(\mathcal{H})$ where $\mathcal{H}$ contains $P_D=SL(2,\C)$ massive singlet states in $L^2(\R^{1,3},\mathcal{W})$ where $\R^{1,3}$ parametrizes position-type or momentum-type coherent states defined in {\S\ref{Poincare}}.} $(\square+m^2)$ for $m\neq0$. According to \emph{(\ref{functional power})} the inverse operator is (for localization $\lambda_{\mathfrak{h}_{\mathrm{U}}}$)
\begin{equation}
 \langle i|(\square+m^2)^{-\alpha}_\Gamma|j\rangle
=\frac{1}{\Gamma(\alpha)}\int_{i\R_+\cup i\R_-} \langle i|e^{-
(\square+m^2)\tau}\,\tau^{\alpha}|j\rangle\;d\log(\tau)\;.
\end{equation}

This representation is to be understood as an operator identity insofar as specific realizations depend on the state vectors/matrix elements under consideration. For example, for position-type states $\psi_{q_a}$, $\psi_{q_b}$ with initial position $q_a,\,q_b\in\R^{1,3}$, appendix \emph{\ref{propagator}} calculates
\begin{equation}
\langle\psi_{q_b}| e^{-(\square+m^2)(\Delta t)}\,\psi_{q_a}\rangle
=e^{-S(q_a,t_a;q_b;t_b)}
=s^{-1}\left(s\Delta t\right)^{-2}e^{-\frac{\Delta t}{s}\left(\frac{\pi(\Delta q)^2}{(\Delta t)^2}-m^2\right)}
\end{equation}
where $\Delta q:=q_b-q_a$, $s^2=4\pi$, and $s\Delta t:=\pm\sqrt{4\pi}(t_b-t_a)$ which has units of $[m]^{-2}$. For $\alpha=1$ this yields a Bessel function presentation of the position-to-position elementary kernel/Green's function/propagator
\begin{eqnarray}
 \langle\psi_{q_b}|(\square+m^2)^{-1}_\Gamma\,\psi_{q_a}\rangle\notag\\
&&\hspace{-1.25in}=\left(\int_{0}^{i\infty}+\int_{0}^{-i\infty}\right)e^{-S(q_a,t_a;q_b,t_b)}\, \tau\,d\log (\tau)\notag\\
&&\hspace{-1.25in}=\left(\int_{0}^{\infty}+\int_{0}^{-\infty}\right)
i s^{-1}\left(s\,|\tau|\right)^{-2}e^{-\frac{1}{s}\left(\frac{\pi(q_b-q_a)^2}{|\tau|}
-m^2|\tau|\right)}\;d|\tau|\notag\\
&&\hspace{-1.25in}=
\frac{-i m}{4\pi^2\sqrt{-(q_b-q_a)^{2}}}K_{(1)}(m\sqrt{-(q_b-q_a)^{2}})
\;,\;\;\;\;\;(q_b-q_a)^{2}<0\,\,,\,\,m^{2}>0\notag\\
\end{eqnarray}
where we used $s\Delta t=-\sqrt{4\pi}(t_b-t_a)\equiv -\sqrt{4\pi}\,\tau$ in the second line.  Given $m^2>0$, the integral over $i\R_-$ does not converge for time-like separations $(q_b-q_a)^{2}<0$. If instead we choose $s\Delta t:=-\sqrt{4\pi}(t_a-t_b)=\sqrt{4\pi}\,\tau$, then the integral over $i\R_-$ gives the propagator while $i\R_+$ fails to converge (given $m^2>0$).

On the other hand, for state vectors $\psi_{p_a}$, $\psi_{p_b}$ with initial $4$-momenta $p_a$, $p_b$, the effective action is the Fourier transform $S(p_a,t_a;p_b,t_b)=\mathcal{F}(S(x_a,t_a;x_b,t_b))$ (see appendix \emph{\ref{propagator}}). This gives the momentum-to-momentum realization
\begin{eqnarray}
 \langle\psi_{p_b}|(\square+m^2)^{-1}_\Gamma\,\psi_{p_a}\rangle
 &=&\int_{i\R_+\cup i\R_-}\langle\psi_{p_b}| e^{-(\square+m^2)(\tau)}\,\psi_{p_a}\rangle \tau\;d\log (\tau)\notag\\
&&\hspace{-1.25in}=\left(\int_{0}^{i\infty}+\int_{0}^{-i\infty}\right)e^{-S(p_a,t_a;p_b,t_b)}\tau
\;d\log (\tau)\notag\\
&&\hspace{-1.25in}=\left(\int_{0}^{i\infty}+\int_{0}^{-i\infty}\right)
is^{-1}(s\tau)^{-2}(s\tau)^2e^{-\frac{\tau}{4{\pi}s}(p_b^2s^2-4\pi m^2)}\tau\;d\log (\tau)\notag\\
\end{eqnarray}
with $p_a=p_b$ and $s\tau=-\sqrt{4\pi}\,\tau$. In general, a functional inverse power such as $(\square+m^2)^{-1}_\Gamma$ comprises a principle value plus a delta function (\emph{\cite[def. $5.5$]{LA3}} and \emph{\cite[$\S$ $4.1$]{LA2}}). Consequently
\begin{equation}
\langle\psi_{p_b}|(\square+m^2)^{-1}_\Gamma\,\psi_{p_a}\rangle
=\frac{i}{p_b^2-m^2}\pm \delta(p_b^2-m^2)\;\;\;\;\mathrm{for}\;i\R_\pm\;.
\end{equation}
Choosing instead $s\tau=\sqrt{4\pi}\,\tau$ exchanges $i\R_+\leftrightarrow i\R_-$. Looking ahead to section \emph{\ref{Poincare}}, the momentum parametrizing space is obtained by an embedding $\R^{1,3}\hookrightarrow\C^3$ so the conditions $\Im(p_b^2-m^2)<0$ for $i\R_+$ and $\Im(p_b^2-m^2)>0$ for $i\R_-$ naturally incorporate the Feynman prescription for the Fourier transformed propagator.\footnote{\label{time inversion}Note that unitarity of $\rho$ in functional Mellin implies  the position-to-position as well as the momentum-to-momentum $i\R_\pm$ kernels are exchanged under $\tau\rightarrow\tau^{-1}=\tau^\dag$ as one can verify by direct calculation. In this sense, time inversion (and not time reversal) looks like conjugate propagation. This is a reflection of the multiplicative-group nature of the evolution time interval and suggests $\tau$ should \emph{not} be identified with the `time' direction of $\R^{1,3}$ --- hence motivating our prior interpretation of $\tau$ as a `time' interval. It also suggests extending to complex one-parameter subgroups with localization $\lambda_{\mathfrak{h}_{\mathrm{U}}}:\phi_{\,\mathfrak{h}_{\mathrm{U}}}(\C)\rightarrow \C^\times$ since $G_D$ may be a \emph{complex} subgroup of $G^\C$: The nature of $\tau$ in this case is much more interesting and explored in appx. \ref{complex time}.}
\end{example}
The purpose of presenting this rather trivial and already well-known example is two-fold: First, it provides an explicit functional Mellin representation of a standard (inverse) operator. Second and more importantly, it illustrates the use of QM functional integrals to calculate propagators that are normally calculated in the context of QFT.

\subsection{ Dynamics}
A first remark is that nontrivial dynamics is possible only for non-commutative $\mathfrak{C}^\ast\equiv L_B(\mathcal{H})$ and non-abelian $G^\C$ (which means functional Mellin will be a $\ast$-representation only when $\alpha=1$ and $\rho$ is unitary). An important property of $\mathbf{F}_{\mathbb{S}}(G^\C)$ necessary for consistency is that it contains a copy of $G_D$;
\begin{proposition}\label{group injection}\emph{(\cite{W} prop. 2.34)}
Let $U(\mathbf{F}_{\mathbb{S}}(G^\C)))$ denote the group of unitaries  in $\mathbf{F}_{\mathbb{S}}(G^\C)$. Define $i_{G^\C}:G^\C\rightarrow U(\mathbf{F}_{\mathbb{S}}(G^\C))$ by $(i_{G^\C}(g_o)\ast\mathrm{F})(g):=Inn(\rho(g^{-1}_o))\mathrm{F}(gg_o)
=\rho(g_o)\mathrm{F}(g)$ with $g_o,g\in G^\C$. Then the map $i_{G^\C}$ is a faithful, unitary-valued homomorphism.
\end{proposition}

\emph{Proof}: Clearly $i_{G^\C}$ is injective because $\rho$ is injective according to Definition \ref{Mellin def.}. Since $\rho$ is unitary and $\alpha=1$ we have
\begin{eqnarray}
\mathit{\Pi}^{(1)}_\lambda(i_{G^\C}(g_o)\ast\mathrm{F})
&=&\left.\int_{G^\C}\rho(g_o)\mathrm{F}(gg^{1}g_o)\rho(g_o^{-1})\;\mathcal{D}_\lambda g\right|_{\alpha=1}\notag\\
&=&\left.\int_{G^\C}\rho(g_o)\mathrm{F}(gg^{1})\rho(g_o)\rho(g_o^{-1})\;\mathcal{D}_\lambda g\right|_{\alpha=1}\notag\\
&=&\rho(g_o)\mathit{\Pi}^{(1)}_\lambda(\mathrm{F})\;.
\end{eqnarray}
 In particular, $\mathit{\Pi}^{(1)}_\lambda(i_{G^\C}(g_o)\ast\mathrm{E}^{-\mathrm{Id}})
 =\rho(g_o)\mathit{\Pi}^{(1)}_\lambda(\mathrm{E}^{-\mathrm{Id}})=\rho(g_o)$ and their algebraic product gives $\mathit{\Pi}^{(1)}_\lambda(i_{G^\C}(g_o)\ast i_{G^\C}(\widetilde{g_o}))=\rho(g_o)\rho(\widetilde{g_o})=\rho(g_o\widetilde{g_o})
 =\mathit{\Pi}^{(1)}_\lambda(i_{G^\C}(g_o\widetilde{g_o}))$.\footnote{Despite appearances, our map $i_{G^\C}$ coincides with $i_G$ of \cite{W}. They are defined differently because we use equivariant functionals $\mathrm{F}$.}
$\QED$
Hence, group elements under $i_{G^\C}$ are observables, and $\mathit{\Pi}^{(1)}_\lambda(\mathbf{F}_{\mathbb{S}}(G^\C))$ contains the associated operators. In particular, this holds for the dynamical group $G_D$.

Now, according to assumption 2.3, the dynamics of a closed quantum system are generated by an inner automorphism $\mathrm{F}\mapsto \mathrm{Inn}(\mathrm{g}(t))\mathrm{F}= \mathrm{g}^{-1}(t)\ast\mathrm{F}\ast\mathrm{g}(t)=:\mathrm{F}(t)$ where $\mathrm{F}\in\mathbf{F}_{\mathbb{S}}(G^\C)$ and $\mathrm{g}(t):=i_{G^\C}(g(t))\in U(\mathbf{F}_{\mathbb{S}}(G^\C))$ with $g\in G_D$. Here $t\in\R$ parametrizes a continuous curve in $G^\C$, and we will call it dynamical-time.  If $\mathrm{g}(t)$ is differentiable,
\begin{equation}\label{ad map}
\frac{d}{dt}\mathrm{F}(t):=\frac{d}{dt}\mathrm{Inn}(\mathrm{g}(t))\,\mathrm{F}
=i\mathrm{Inn}'_{\mathrm{g}(t)}(\mathfrak{g}_{\mathrm{g}(t)})\mathrm{F}
=:\mathrm{inn}(i\mathfrak{g}_{\mathrm{g}(t)})\mathrm{F}
\end{equation}
where $d\mathrm{g}(t)/dt=i\mathfrak{g}_{\mathrm{g}(t)}
=iL'_{\mathrm{g}(t)}\mathfrak{g}_{\mathrm{g}(0)}$ and $\mathrm{g}(0)=\mathrm{Id}$.  In other words, $\mathbf{F}_{\mathbb{S}}(G^\C)$  models the Lie bracket structure ostensibly possessed by $\mathfrak{A}_L$ through the derivative of the $\mathrm{Inn}$ action. This yields corresponding inner automorphism actions of ${G_D}_\lambda$ and ${\mathfrak{G}_D}_\lambda$ on $L_B(\mathcal{H})$ given by
\begin{eqnarray}
\mathit{\Pi}_\lambda^{(1)}\left(\mathrm{Inn}(\mathrm{g}(t))\mathrm{F}\right)
&=&\mathit{\Pi}_\lambda^{(1)}\left(\mathrm{g}^{-1}(t)\ast\mathrm{F}\ast\mathrm{g}(t)\right)\notag\\
&= & \mathit{\Pi}_\lambda^{(1)}\left(\mathrm{g}^{-1}(t)\right)
\,\mathit{\Pi}_\lambda^{(1)}\left(\mathrm{F}\right)
\,\mathit{\Pi}_\lambda^{(1)}\left(\mathrm{g}(t)\right)\notag\\
&=&Inn(g(t))\mathit{\Pi}_\lambda^{(1)}\left(\mathrm{F}\right)
\end{eqnarray}
 and
\begin{equation}
\mathit{\Pi}^{(1)}_\lambda(\mathrm{inn}(i\mathfrak{g}_{\mathrm{g}(t)})\mathrm{F})
=inn(i\rho'(\mathfrak{g}_{\mathrm{g}(t)}))\mathit{\Pi}^{(1)}_\lambda(\mathrm{F})
=i[\rho'(\mathfrak{g}_{\mathrm{g}(t)}),{F}]\;.
\end{equation}

The next task is to construct the evolution operator. Assume a skew-adjoint, possibly dynamical-time-dependent evolution generator of the form $\mathit{\Pi}_\lambda^{(1)}\left(\mathrm{H}(t)\right)
={H}(t)$. Using Magnus' expansion\cite{MA}, the evolution-generating observable $\mathrm{U}_{t_b,t_a}:=\mathcal{T}\left\{\mathrm{E}^{-\int_{t_a}^{t_b}\mathrm{H}(t)\,dt}\right\}$ can be represented as
\begin{equation}
\mathit{\Pi}^{(1)}_\lambda(\mathrm{U}_{t_b,t_a})=U_{t_b,t_a}
=\mathcal{T}\left\{e^{-\int_{t_a}^{t_b}{{H}}(t)\,dt}\right\}
=e^{-\int_{t_a}^{t_b}\widetilde{{H}}(t)\,dt}
\end{equation}
such that
\begin{equation}
\frac{d\widetilde{{H}}(t)}{dt}
=\sum_{n=0}^\infty\frac{B_n}{n!}\,{inn}^n\left({H}(t)\right)
\widetilde{{H}}(t)
\end{equation}
where $B_n$ are Bernoulli numbers. The $n$-th power adjoint map ${inn}^n\left({H}(t)\right)$ is defined by recursion starting with ${inn}^0\left({H}(t)\right)
\widetilde{{H}}(t):=\widetilde{{H}}(t)$ at the first level $n=0$ and then recursively by
${inn}^n\left(\mathrm{H}(t)\right)
\widetilde{{H}}(t):={inn}^1\left({H}(t)\right)
{inn}^{n-1}\left({H}(t)\right)
\widetilde{{H}}(t)$.
This yields a representation of the Heisenberg equation
\begin{equation}
\mathit{\Pi}_\lambda^{(1)}\left(\frac{d\mathrm{F}(t)}{dt}\right)
={inn}(\widetilde{H}(t))F
=\left[\widetilde{H}(t),F\right]=\frac{dF(t)}{dt}\;.
\end{equation}

Meanwhile, with $t_a\equiv 0$ and $t_b\equiv t$, the observable $\mathrm{U}_t$ generates an inner automorphism on the group $G_\lambda^\C$ represented by
\begin{equation}
\mathit{\Pi}_\lambda^{(1)}(g(t))=\rho(g(t))
:={U}_t^{-1}\rho(g(0)){U}_t
\end{equation}
whose derivative mapping defines the adjoint representations of ${G_D}_\lambda$ and ${\mathfrak{G}_D}_\lambda$ on $L(\mathcal{H})$.
In particular, for $U(\mathbf{F}_{\mathbb{S}}(G^\C)))\ni \mathrm{g}(t)=\mathrm{E}^{-\mathfrak{g}(t)}$ with $\mathfrak{g}(t)\in\mathfrak{G}_\lambda$,
\begin{equation}
\rho'(\mathfrak{g}(t))
={U}^{-1}_t\,\rho'(\mathfrak{g}(0)){U}_t\;,\;\;\;\;\;\;
\frac{d\rho'(\mathfrak{g}(t))}{dt}
=\left[\widetilde{H}(t),\rho'(\mathfrak{g}(0))\right]\;.
\end{equation}
Note $\rho'(\mathfrak{g}(t))\in L(\mathcal{H})$ is not necessarily bounded. Nevertheless, if $\mathfrak{g}(t)$ is skew-adjoint, it is observable and possesses a dynamical-time-dependent spectrum that represents a changing kinematical description in the sense that the parabolic decomposition of $\mathfrak{G}_\lambda^\C$ and the consequent induced representation $\rho$  are dynamical-time dependent in general. Hence the spectrum of $\rho({P_D}_\lambda)$ is not necessarily constant. However, the parabolic decomposition is stable provided $\mathfrak{P}_D$ is invariant under $inn(\widetilde{H}(t))$.

It is convenient to formulate transition amplitudes using Hilbert $\mathcal{H}=L^2(X,\mathcal{W})$. Write ${\psi}_t:=U_t{\psi}$ where $\psi\in L^2(X,\mathcal{W})$. Remind that $U_t$ depends implicitly on $\lambda$. As usual, unitarity supplies the connection between the Heisenberg and Schr\"{o}dinger pictures
\begin{equation}
\langle{\phi}|\mathit{\Pi}^{(1)}_{\lambda}(\mathrm{F}(t)){\psi}\rangle
=\langle{\phi}|U_t^{-1}\,\mathit{\Pi}^{(1)}_\lambda(\mathrm{F})\,U_t|{\psi}\rangle
=\langle{\phi}_t|\mathit{\Pi}^{(1)}_\lambda(\mathrm{F})|{\psi}_t\rangle
\;.\notag
\end{equation}
Define the $\ast$-homomorphism $\pi_x:L_B(L^2(X,\mathcal{W}))\rightarrow L_B(\mathcal{W}_x)$ by
\begin{equation}\label{pi_x}
\pi_x(\mathit{\Pi}_\lambda^{(1)}(\mathrm{F}))\psi(x):=
(\mathit{\Pi}_\lambda^{(1)}(\mathrm{F})\psi)(x)\;\;\forall x\in X\;.
\end{equation}
Explicitly, in a local trivialization $\pi_x(\mathit{\Pi}_\lambda^{(1)}(\mathrm{F}))\psi(x)=(x,\Bold{v}_{\mathrm{F}})$
where $(x,\Bold{v})$ is the representative of $\psi(x)$ and $\Bold{v}_{\mathrm{F}}=(\mathit{\Pi}_\lambda^{(1)}(\mathrm{F})\breve{\psi})(s_i(x))$ relative to a canonical section.
Writing $\psi_t(x)=\psi(x;t)$, the evolution-time-dependent transition amplitude in the Heisenberg and Schr\"{o}dinger pictures can be expressed as
\begin{eqnarray}
\langle{\psi}_b|\mathit{\Pi}^{(1)}_{\lambda}(\mathrm{F}(t)){\psi}_a\rangle_{\mathcal{H}_D}
&=&\int_{X_D}({\psi}_b(x)|\pi_x(\mathit{\Pi}_\lambda^{(1)}\mathrm{F}(t))
{\psi}_a(x))_{\mathcal{W}_x}\;d\mu_{P_D}(x)\notag\\
&=&\int_{X_D}({\psi}_b({x};t)|\pi_x(\mathit{\Pi}_\lambda^{(1)}\mathrm{F})
{\psi}_a({x};t))_{\mathcal{W}_{x}}\;d\mu_{P_D}(x)\;.\notag\\
\end{eqnarray}
 When $\int_{t_a}^{t_b}\widetilde{H}(t)\,dt\sim H(t_b-t_a)$ we identify $t\equiv\tau$, but we emphasize that even in this case $\tau$ is not \emph{a priori} identified with a subset of the spectrum of $\rho({X_D})$.

The above expectation values are associated with dynamics and must therefore be restricted to $\mathcal{H}_D$. Assuming that ${G_D}_\lambda$ is connected, simply connected, and $H^2({G_D}_\lambda,\R)\cong0$, then we are in business in the sense that the $S$-matrix will enjoy the symmetries associated with $G_D$ and $P_D$.(see \S\ref{preliminaries})

\section{Reality check}
It is fitting that we check the proposed quantization procedure on non-relativistic and relativistic quantum mechanics. Our exposition in this section is an outline at a fairly superficial --- but hopefully adequate --- level to relate pertinent objects to their standard counterparts. This exercise illustrates the proposed framework in familiar contexts, provides some functional Mellin tools, and suggests some unconventional interpretations.

\subsection {Non-relativistic QM}
This subsection is not so much a check on the formalism as a demonstration: As we have stated, under suitable conditions functional Mellin with $\alpha=1$ coincides with crossed products\cite{W}, and the crossed product machinery has been substantially developed and checked in the non-relativistic QM context. (Indeed, crossed products were historically constructed precisely to access the $C^\ast$-algebraic foundations of non-relativistic QM.) Here we outline only the simplest case of a scalar particle.

\subsubsection{Induced representation}
Non-relativistic quantum mechanics of a scalar particle \emph{relative to a fixed Galilean reference frame in $\R^3$} is governed by the Heisenberg algebra $\mathfrak{H}^3:=\mathrm{span}_\R\{\mathfrak{h}\}
\,\oplus\,\mathrm{span}_\R\{\mathfrak{e}_{a},\mathfrak{e}^\dag_{a}\}$ with brackets
\begin{equation}\label{Heisenberg}
[\mathfrak{e}_a,\mathfrak{e}^\dag_b]
=\delta_{a,b}\,\mathfrak{h};\;\;\;[\mathfrak{e}_a,\mathfrak{e}_b]=0
;\;\;\;[\mathfrak{e}^\dag_a,\mathfrak{e}^\dag_b]=0
;\;\;\;[\mathfrak{h},\mathfrak{e}_a]=[\mathfrak{h},\mathfrak{e}^\dag_a]=0
\end{equation}
where $\mathfrak{e}_{-a}:=\mathfrak{e}_a^\dag$ is conjugate to $\mathfrak{e}_a$ and $a,b\in\{1,2,3\}$.
The algebra possess a triangular decomposition, and it can be realized as creation/annihilation operators on a bosonic Fock space in the usual manner with excitations interpreted as energy levels.

The exponentiated algebra $\mathfrak{H}^3$ is the Heisenberg group $H^3$ whose manifold structure is $\mathbb{M}\,[H^3]\cong\R^6\times\,\R$. But the same algebra also characterizes the reduced Heisenberg group $\overline{H^3}$ with underlying manifold $\mathbb{M}\,[\overline{H^3}]\cong\R^6\times U(1)$. There doesn't seems to be observable physics that differentiates between the two groups so let's take $G^\C_{\lambda}=(\overline{H^3})^\C$ as our locally compact topological group and assume the dynamical group $G_D$ is the associated real group $\overline{H^3}$.

Observing that $\mathfrak{u}(1)$ is the maximal compact subalgebra, we choose the parabolic subalgebra $\mathfrak{P}^\C=\mathfrak{E}\oplus\mathfrak{u}(1):=\mathrm{span}_\C\{\mathfrak{e}_{a}\}\oplus\mathfrak{u}(1)$. Then the induced unitary representations labeled by a phase $\theta$ are determined by equivariant maps $\breve{\psi}\in L^2((\overline{H^3})^\C,\mathcal{V}_{(\theta)})$ where $\theta\in(0,2\pi]$ and $\mathcal{V}_{(\theta)}$ is a one-dimensional, $\mathfrak{P}^\C$-invariant subspace of the highest-weight module generated by $\overline{H^3}$. For any given choice of $\theta$, it is well known that these are all irreducible and unitarily equivalent.

Consequently, fix the representation with a fiducial phase choice $\theta_0\equiv\hbar$ and construct the line bundle $\mathcal{W}_{\mathcal{V}}=(\mathcal{W},\C^3,pr, \mathcal{V}_{(\hbar)}, P^\C)$ where $P^\C=(\overline{H^3})^\C/\C^3$. This line bundle is the vector bundle associated to the principal bundle $((\overline{H^3})^\C,\C^3,\breve{pr},P^\C)$. Given a local trivialization and a canonical section on the principal bundle, an element $\psi\in\Gamma(\C^3,\mathcal{W})$ with appropriate normalization and constraints/boundary conditions can be identified with $\breve{\psi}$, and we can construct the quantum Hilbert space of state vectors $\mathcal{H}:=L^2(\C^3,\mathcal{W})$ with a suitable quasi-invariant measure $\mu_{P^\C}$. The various instantiations of this representation and its relation to the bosonic Fock space realization is a standard text-book lesson that does not require repeating. Suffice it to say that the holomorphic wave function $\psi(z)\in\mathcal{W}$ is interpreted as the probability amplitude associated with a particle at the point $z\in\C^3$ subject to certain constraints and/or boundary conditions.

The required covariance of $\psi$ with respect to $P_D:=\overline{H^3}/\R^3 $ adds one more feature: The right action of $\mathfrak{e}_a$ on $\C^3$ generates translation so equivalence classes of physical state vectors are distinguished by three labels $(\mathrm{p}_1,\mathrm{p}_2,\mathrm{p}_3)$. Restrict to a real slice $\R^3\subset\C^3$ and identify this $\R^3$ as the parametrizing space of position. Then the equivalence class labels $(\mathrm{p}_1,\mathrm{p}_2,\mathrm{p}_3)=:\mathbf{p}$ can be interpreted as a momentum, and physical state vectors can be represented by $[\psi^{\hbar}]_{\mathbf{p}}$ with $\mathbf{p}\in\R^3$. It is natural to identify this state vector with the probability amplitude of a non-relativistic particle of momentum $\mathbf{p}$ represented by the position wave function $\psi_{\mathbf{p}}(\mathbf{x})$.\footnote{Alternatively, identify the real slice $\R^3$ with the parametrizing space of momentum. Then $[\psi^{\hbar}]_{\mathbf{x}}$ gives rise to a momentum wave function $\psi_\mathbf{x}(\mathbf{p})$ that is Fourier dual to $\psi_{\mathbf{p}}(\mathbf{x})$.}

\subsubsection{QM algebra and dynamics}
By previous hypothesis, a quantum system is modeled on the $C^\ast$-algebra of integrable functionals $\mathbf{F}_{\mathbb{S}}(G^C)$ on some topological group $G^\C$. In this example, for a specific system determined by a particular choice of $\lambda$, the topological group reduces to the locally compact topological group $(\overline{H^3})^\C$. Then, relevant operators along with their trace, log, and determinant can be represented by Mellin functional integrals $\mathcal{M}_\lambda[\mathrm{F};\alpha]$ as long as
$\mathrm{F}$ and its associated function $f\in L^1((\overline{H^3})^\C,L_B(\mathcal{H}))$ are Mellin integrable with fundamental strip $\mathbb{S}\supseteq\langle0,1\rangle$. Evolution is governed by unitary elements $\mathrm{E}^{-\mathrm{H}}\in G_D\subseteq UM(\mathbf{F}_{\mathbb{S}}(G^\C))$ where the observable $\mathrm{H}(t)$  generates dynamics. Its associated operator $(\mathit{\Pi}^{(1)}_\lambda)'(\mathrm{H}(t))$ springs from the reduced Heisenberg group, and in this sense $\overline{H}^3$ is the {dynamical} group.

To finish the description of the dynamical system requires specification of the physical Hilbert space $\mathcal{H}_D$ representing ${G_D}_\lambda=\overline{H}^3$.  Since $H^2(\overline{H}^3,\R)\cong0$, the ordinary UR $\rho:(\overline{H^3})^\C\rightarrow L_B(\mathcal{H})$ just restricts and $\mathcal{H}_D$ furnishes sub-representations labeled by $(\theta,\mathbf{p})$.  Appealing to the geometric quantization program suggests that restricting to sub-representations in this case implies a choice of a real/holomorphic polarization on $T\C^3$.

\begin{remark}\label{classical phase space}
Consider a state vector  that injectively maps a neighborhood $U_i\subset \R^3\rightarrow \mathcal{W}$. In particular, in a local trivialization a constant section mapping $\mathbf{x}\mapsto(\mathbf{x},{v}_{w_+})$ for all $\mathbf{x}\in U_i$ can be interpreted as the ground state $\psi_0$. Then under evolution by some unitary operator $U_\tau$, the new state $U_\tau\psi_0$ will map $U_i\subset\R^3$ to some (generally) nontrivial subspace in $\mathcal{W}$. In other words, $\mathbb{D}_{\mathbf{x}}(t)
:=\left(\bar{\varrho}'(\mathfrak{E}(t)+i\mathfrak{E}^\dag(t))\,\psi_0\right)(\mathbf{x})$  is a dynamical-time-dependent embedding $U_i\hookrightarrow\mathcal{W}$ with $\mathrm{dim}_\R(\mathbb{D}_{\mathbf{x}}(t))\leq3$ (given suitable analytic/differentiable conditions on $U_\tau\psi_0(\mathbf{x})$).

If we eschew the conventional probability-amplitude interpretation of $\psi_{\mathbf{p}}(\mathbf{x})\in\Gamma(\R^3,\mathcal{W})$, the evolution of the ground state generated by $(\mathfrak{E}(t)+i\mathfrak{E}^\dag(t))$ can be interpreted as giving rise to a configuration manifold $\mathbb{Q}(t)$ whose tangent space at $\mathbf{x}$ is $T_\mathbf{x}\mathbb{Q}(t)=\mathbb{D}_{\mathbf{x}}(t)$ with Lie bracket coming from $[\mathfrak{E}(t),\mathfrak{E}^\dag(t)]$. Further, for $\mathbb{P}_{\mathbf{x}}(t)
:=\left(\bar{\varrho}'(\mathfrak{E}(t)-i\mathfrak{E}^\dag(t))\,\psi_0\right)(\mathbf{x})$ a topological space, the cotangent bundle $T^\ast\mathbb{Q}(t):=\mathbb{Q}(t)\times\mathbb{P}(t)$ can be thought of as a dynamical-time-dependent phase space associated with  $\psi_0$. If, in particular, the Hamiltonian has the dynamical-time-independent form $\mathrm{H}=1/2\mathfrak{p}^2+\mathrm{V}(\mathfrak{q})$ for a suitable potential $\mathrm{V}(\mathfrak{q})$ where $\mathfrak{p}\in(\mathfrak{E}-i\mathfrak{E}^\dag)$ and $\mathfrak{q}\in(\mathfrak{E}+i\mathfrak{E}^\dag)$, then the embedding is static since $\langle\psi_0|\rho'(d\mathfrak{q}/dt)\psi_0\rangle=\langle\psi_0|\rho'(\mathfrak{p})\psi_0\rangle=0$, and hence $\mathbb{Q}\cong\R^3$.

Now introduce a formal reproducing kernel on $T^\ast\mathbb{Q}$ by $K(\mathbf{p},\mathbf{q})
:=\langle\psi_0|e^{\sum_{a}p_a\mathfrak{p}_a}
e^{\sum_{b}q_b\mathfrak{q}_b}\,\psi_0\rangle$ with $(\mathbf{p},\mathbf{q})\in T^\ast\mathbb{Q}$. Using BCH and the fact that $\mathfrak{h}$ is a central element, the normalized kernel can be written
\begin{equation}
\widetilde{K}(\mathbf{p},\mathbf{q})
:=\frac{\langle\psi_0|e^{\sum_{b}(iq_b)\mathfrak{e}^\dag_b}\,
e^{\frac{1}{2}(iq_b)p_a\delta_{a,b}\mathfrak{h}}\,
e^{\sum_{a}p_a\mathfrak{e}_a}\,\psi_0\rangle}{\langle\psi_0|\psi_0\rangle}
=e^{i\hbar\,\mathbf{p}\cdot\mathbf{q}}\;.
\end{equation}
Then by Fourier, $\widetilde{K}(\mathbf{q},\mathbf{q}')=\delta(\mathbf{q}'-\mathbf{q})$ and $\widetilde{K}(\mathbf{p},\mathbf{p}')=\delta(\mathbf{p}-\mathbf{p}')$. These, of course, are elementary and well known.
In this picture, the CS ``wave function" $\psi_{\mathbf{p}}(\mathbf{x})$ is a subordinate notion to the ``expected" cotangent bundle $T^\ast\mathbb{Q}(t)$, and the evolution of a quantum state  $[\psi^{\hbar}]_{\mathbf{p}}$, say with fixed initial and final position contained in $\mathbb{Q}$, must then be associated with \textbf{all} continuous paths in $T^\ast\mathbb{Q}$ with the specified end-points (due to the vertical automorphism covariance of physical states).

There are two problems with this picture. First, $\mathbb{Q}$ is not Minkowski. Second, $T^\ast\mathbb{Q}$ only describes evolution of a single particle type with given boundary conditions. To be realistic, the quantum Hilbert space should include a suitable sum over all relevant particle types, say for example $[\psi^{\hbar}]_{\mathbf{p},m_0}$ labeled by momentum and rest mass. However, the assumed group $G^\C$ and its observed locally compact ``shadow'' ${\overline{H}^3}$ only allow to describe a single particle type unless we want to consider a direct sum of Hilbert spaces over all particle types labeled by rest mass. The problem with this fix is that ${\overline{H}^3}$, which is supposed to tell all, wouldn't allow for transitions between the single-particle Hilbert spaces. One solution to both problems, of course, is to replace the Heisenberg group with Poincar\'{e} to which we now turn.
\end{remark}

\subsection{Poincar\'{e} QM}\label{Poincare}
Since the Poincar\'{e} algebra doesn't have a triangular decomposition and hence does not conform to Assumption 2.1, let's instead begin with $G^\C_\lambda=SO(5,\C)$. We will recover Poincar\'{e} QM via the standard group contraction of anti deSitter (AdS) $SO(3,2)\rightarrow ISO(3,1)$.

Begin with the group algebra
\begin{equation}
[\mathfrak{e}_{ij},\mathfrak{e}_{kl}]
=\eta_{jk}\mathfrak{e}_{il}+\eta_{il}\mathfrak{e}_{jk}
-\eta_{jl}\mathfrak{e}_{ik}-\eta_{ik}\mathfrak{e}_{jl}
\end{equation}
where $\mathfrak{e}_{ji}=-\mathfrak{e}_{ij}$ with $i\in\{0,1,2,3,4\}$ and $\eta=\mathrm{diag}(-1,1,1,1,-1)$ is the inner product on $\mathfrak{so}(5)$ that induces a metric on the underlying group manifold with signature $(-,+,+,+,-)$. Its maximal compact subalgebra is $\mathfrak{so}(3)\oplus \mathfrak{so}(2)$ generated by, say, $\mathfrak{h}:=i\mathfrak{e}_{04}$ and $\mathfrak{j}_{ab}:=i\mathfrak{e}_{ab}$ where $a,b\in\{1,2,3\}$. Accordingly,  induced URs relevant to the eventual contraction employ the algebra decomposition
\begin{eqnarray}\label{first decomposition}
i[\mathfrak{j}_{ab},\mathfrak{j}_{cd}]
&=&\eta_{bc}\mathfrak{j}_{ad}+\eta_{ad}\mathfrak{j}_{bc}
-\eta_{bd}\mathfrak{j}_{ac}-\eta_{ac}\mathfrak{j}_{bd}\notag\\
i[\mathfrak{z}^\pm_{a},\mathfrak{j}_{bc}]&=&\pm\eta_{ab}\mathfrak{z}^\pm_{c}
\mp\eta_{ac}\mathfrak{z}^\pm_{b}\notag\\
\,[\mathfrak{z}^\pm_{a},\mathfrak{h}]&=& \mp \mathfrak{z}^\pm_{a}\notag\\
\,[\mathfrak{z}^+_{a},\mathfrak{z}^-_{b}]&=&-2(\eta_{ab}\mathfrak{h}+i\mathfrak{j}_{ab})\notag\\
\,[\mathfrak{z}^\pm_{a},\mathfrak{z}^\pm_{b}]&=&[\mathfrak{j}_{ab},\mathfrak{h}]=0
\end{eqnarray}
where $\mathfrak{z}^+_a:=(\mathfrak{e}_{0a}+i\mathfrak{e}_{a4})$ and its conjugate $\mathfrak{z}^-_a:=(\mathfrak{z}^+_a)^\dag=(\mathfrak{e}_{0a}-i\mathfrak{e}_{a4})$.  Group contraction is implemented through $\mathfrak{e}_{04}\rightarrow R^{-1}\mathfrak{e}_{04}$  and $\mathfrak{e}_{a4}\rightarrow R^{-1}\mathfrak{e}_{a4}$ with $R$ the AdS radius.

It's easier to visualize the contraction using the alternative decomposition
\begin{eqnarray}\label{second decomposition}
i[\mathfrak{j}_{\mu\nu},\mathfrak{j}_{\sigma\rho}]
&=&\eta_{\nu\sigma}\mathfrak{j}_{\mu\rho}+\eta_{\mu\rho}\mathfrak{j}_{\nu\sigma}
-\eta_{\nu\rho}\mathfrak{j}_{\mu\sigma}-\eta_{\mu\sigma}\mathfrak{j}_{\nu\rho}\notag\\
i[\mathfrak{z}_{\mu},\mathfrak{j}_{\nu\sigma}]&=&\eta_{\mu\nu}\mathfrak{z}_{\sigma}
-\eta_{\mu\sigma}\mathfrak{z}_{\nu}\notag\\
 \,[\mathfrak{z}_{\mu},\mathfrak{z}_{\nu}]&=&\mathfrak{j}_{\mu\nu}
\end{eqnarray}
where $\mu\in\{0,1,2,3\}$ and $\mathfrak{z}_\mu:=\mathfrak{e}_{\mu 4}$. The contraction to Poincar\'{e} is clearly evident in this decomposition. And since the $\mathfrak{z}_\mu$ mutually commute when $R\rightarrow\infty$, they serve to parametrize relevant states via induced representations in this limit. This decomposition isn't triangular, but the machinery of induced representations still works in this case.

It is useful to compare representations induced by both algebra decompositions. Turn first to decomposition (\ref{second decomposition}). Construct the principal bundle $(SO(5,\C),Z,\breve{pr},SL(2,\C))$ and its associated Whitney bundle $(\mathcal{W},Z,pr,\mathcal{W}_{(\Bold{\mu})},SL(2,\C))$ derived from finite-dimensional, highest-weight UIRs of $\mathfrak{su}(2)$. The coset space $Z=SO(5,\C)/SL(2,\C)$ is generated by $\mathfrak{z}_\mu$ and will contract to $\C^{4}$ when the AdS radius $R\rightarrow\infty$. In this limit, {global} sections $\psi\in L^2(Z,\mathcal{W})$ can be defined; precisely because $[\mathfrak{z}_{\mu},\mathfrak{z}_{\nu}]\stackrel{R\rightarrow\infty}{=}0$ (which renders a trivial bundle since the base space is contractible). Post contraction, the dynamical group is $G_D=\R^{1,3}\rtimes SL(2,\C)$,\footnote{Alternatively, one can imagine phase space somehow depends on $R$ and group contraction corresponds to a certain region in phase space determined by the dynamics.} its maximal compact subalgebra is $\mathfrak{su}(2)$, and  $P_D=SL(2,\C)$. Representations are furnished by state vectors $[\psi^{(j,\sigma)}]_{\Bold{b}}\in L^2(Z,\mathcal{W})$. These are equivalence classes labeled by $\mathfrak{su}(2)$ quantum numbers $(j,\sigma)$ and three parameters $\Bold{b}:=(\mathrm{b}_1, \mathrm{b}_2, \mathrm{b}_3)$ coming from Lorentz transformations generated by the right action of $P_D/SU(2)$ on $Z$. The restriction to the physical state subspace requires a choice of embedding $\R^{1,3}\hookrightarrow Z$. Viewing $Z=\C^{4}$ as $\R^{1,3}\oplus i\R^{1,3}$, the equivalence classes can be seen to furnish two groups of complex conjugate representations. The temptation is to attach a momentum-space interpretation to $\R^{1,3}\hookrightarrow Z$ and a particle interpretation to $[\psi^{(j,\sigma)}]_{\Re(\Bold{b})}$. However, this induced representation is not irreducible so the correspondence with physical particles is unclear. In particular, the interpretation of the equivalence-class labels $\Re(\Bold{b})$ is not obvious. One can, of course, use the Poincar\'{e} Casimirs to decompose $L^2(Z,\mathcal{W})$ into a direct sum of irreducible representations and then make particle interpretations. But instead let us go back to the triangular decomposition and employ the notion of coherent states to interpret physical state vectors.

\subsubsection{Coherent states}
Now consider decomposition (\ref{first decomposition}).\footnote{The reader is invited to compare this subsection to standard representation analysis of the $d$-dimensional conformal group.} Before contraction, the relevant parabolic subalgebra is $\mathfrak{P}^\C=\mathrm{span}_\C\{\mathfrak{z}_a^+\}\oplus\mathfrak{so}(3,\C)\oplus\mathfrak{so}(2,\C)$ which induces the coset space $\Z:=SO(5,\C)/P^\C$. Construct the principal bundle $(SO(5,\C),\Z,\breve{pr},P^\C)$ and associated bundle $(\mathcal{W},\Z,pr,\mathcal{W}_{(\Bold{\mu})},P^\C)$ derived from relevant dominant-integral highest weight UIRs of $\mathfrak{so}(3)\oplus\mathfrak{so}(2)$. The inducing UIRs are labeled by $r=(j,\sigma,n)$ with quantum numbers $(j,\sigma)$ from $\mathfrak{so}(3)$ and $n\in\mathbb{Z}$ from $\mathfrak{so}(2)$.  The Hilbert space of states is comprised of equivalence classes
$[\psi^{(r)}]_{\mathbf{z}^+}\in L^2(\Z,\mathcal{W})$ labeled by ${\mathbf{z}^+:=(\mathrm{z}^+_1,\mathrm{z}^+_2,\mathrm{z}^+_3})$ coming from the right action of $\exp\{\mathfrak{z}^+_a\}$ on $\Z$.

 Again, the interpretation of these state vectors is not yet clear since the induced representations are generically reducible and  they are parametrized by $\C^3$. All we know is that they must be either holomorphic or anti-holomorphic functions on $\Z$ because we assumed the symmetry $\mathfrak{G}_-\leftrightarrow\mathfrak{G}_+$ in the algebra decomposition. To facilitate state vector interpretation, we will construct coherent states based on $\mathcal{H}= L^2(\Z,\mathcal{W})$ following Perelemov.(c.f. \cite{DQ,BR})

Recall $g\in SO(5,\C)$ can be viewed as an admissible map $g:\mathcal{W}_{(\Bold{\mu})}\rightarrow pr^{-1}(\mathrm{z})\in\mathcal{W}$. Given a local trivialization $\{U_ i,\varphi_ i\}$ of the Whitney sum bundle $\mathcal{W}$ and a local chart
$\phi:U_ i\rightarrow \Z$, a point in
$pr^{-1}(U_ i)\subset\mathcal{W}$ can be represented on
$\Z\times\mathcal{W}_{(\Bold{\mu})}$ as
\begin{equation}
|\phi(\mathrm{z});\Bold{\mu}):=\left(\exp\left\{\sum_{a}{\mathrm{z}}^\ast_{a}\,\mathfrak{z}^-_{a}\right\}\right)| \Bold{\mu})
\end{equation}
where
$\phi(\mathrm{z})={\mathrm{z}}^\ast_{a}\in \Z$ are coordinates of the point $\mathrm{z}\in U_ i\,$, we used the coset parametrization $g=\exp\{\sum_{a}{\mathrm{z}}^\ast_{a}\,\mathfrak{z}^-_{a}\}\exp\{\mathfrak{p}^\C\}$ with $\mathfrak{p}^\C\in \mathfrak{P}^\C$, and  vector $|\Bold{\mu}):=\exp\{\mathfrak{p}^\C\}|\Bold{v}_{w_+})\in\mathcal{W}_{(\Bold{\mu})}$ issues from a collection of dominant-integral highest weights $\Bold{v}_{w_+}:=({v}_{w_+}^{(r_1)},\ldots,{v}_{w_+}^{(r_n)})$ for UIRs of $SO(3)\times SO(2)$. To simplify notation choose normal coordinates and write
$|\phi(\mathrm{z});\Bold{\mu})\equiv|\mathrm{z}^\ast;\Bold{\mu})$.

With this construction, a physical state vector
$[\psi^{(r)}]_{\mathbf{z}^+}\in\mathcal{H}$ can be modeled locally on
$U_ i\times\mathcal{W}_{(\Bold{\mu})}$  as $({\mathrm{z}};\Bold{\mu}|\psi\rangle=:{\psi}_{\Bold{\mu}}(\mathrm{z}) $ where $\Bold{\mu}$ carries both vector components associated with $r$ and $\mathbf{z}^+$ labels. We call ${\psi}_{\Bold{\mu}}(\mathrm{z})$ a coherent state wave function
or coherent state (CS) for short. A general coherent state is an equivalence class of column vectors comprised of all relevant UIRs of $SO(3)\times SO(2)$ collectively labeled by
$\Bold{\mu}=(\mu^{(r_1, \mathbf{z}^+_1)},\ldots,\mu^{(r_n,\mathbf{z}^+_n)})$. Since
$\mathcal{V}^{(r)}_{w_+}$ furnish UIRs,
${\psi}_{\Bold{\mu}}(\mathrm{z})$ is composed of components
${\psi}_{\Bold{\mu}}(\mathrm{z})
=({\psi}_{\mu^{(r_1,\mathbf{z}^+_1)}}(\mathrm{z}),\ldots,
{\psi}_{\mu^{(r_n,\mathbf{z}^+_n)}}(\mathrm{z}))$ that do not mix --- a kind-of
super selection. Similarly, an operator has a CS realization $\widehat{{O}}\,{\psi}_{\Bold{\mu}}(\mathrm{z}):=({\mathrm{z}};\Bold{\mu}|\,{O}\,\psi\rangle$ with $O\in L_B(\mathcal{H})$.

There are exceptional CS, annihilated by raising operators in $\mathfrak{P}^\C$, that we want to associate with ground states. Define a ground state by $\breve{\psi}_0(g):=\Bold{v}_{\Bold{w}_+}\in\mathcal{W}_{(\Bold{\mu})}
\;\forall g\in G^\C$, and a vacuum state $\breve{\varphi}_0$ to be the ground state of the UIR induced from the trivial representation of $SO(3)\times SO(2)$. For the vacuum, $\mathcal{V}_1^{(0)}\subset\mathcal{W}_{(\Bold{\mu})}$ is one-dimensional; hence $(\rho(p)\breve{\varphi}_0)(g)\propto{v}_{{w}_+}$ for all $p\in P^\C$. Evidently the vacuum in $U_i$ may be identified with the zero section of $\mathcal{W}$ over $U_i$ (up to a fixed constant) and it is appropriate to call $P^\C$ the gauge group and $\mathcal{V}_1^{(0)}$ the gauge-invariant vacuum.  This suggests the definition of the CS model for the
vacuum state vector $\varphi_0\in\mathcal{H}$ as $({\mathrm{z}};\Bold{\mu}|\varphi_0\rangle:=\varphi_{0}(\mathrm{z})
\equiv(\mathrm{z},{v}_{{w}_+})$ for all $\mathrm{z}\in \Z$ with normalization $\langle\varphi_0|\varphi_0\rangle_{\mathcal{H}}=|{v}_{{w}_+}|^2$ where ${v}_{{w}_+}\in\mathcal{V}_1^{(0)}$.

Define a formal reproducing kernel on $U_ i\times\mathcal{W}_{(\Bold{\mu})}$ by
\begin{eqnarray}\label{overlap}
(\Bold{K}(\mathrm{z}',\mathrm{z}^\ast))_{\Bold{\mu}'\,\Bold{\mu}}
&:=&({\mathrm{z}'};\Bold{\mu}'|\mathrm{z}^\ast;\Bold{\mu})
=(\Bold{\mu}'|e^{
\sum_{a}{\mathrm{z}}'_{a}\,\mathfrak{z}^+_{a}}\,e^{
\sum_{b}{\mathrm{z}}^\ast_{b}\,\mathfrak{z}^-_{b}}|\Bold{\mu})
\end{eqnarray}
and the associated
resolution of the identity by
\begin{eqnarray}
{{Id}}
&=&\int_{U_ i}|\mathrm{z}^\ast;\Bold{\mu})\; \mathbf{d\,\Bold{\sigma}}(\mathrm{z})\;(
{\mathrm{z}};\Bold{\mu}|
\end{eqnarray}
where
\begin{equation}
\mathbf{d\,\Bold{\sigma}}(\mathrm{z})
:=\mathcal{N}\,\Bold{K}^{-1}(\mathrm{z},\mathrm{z}^\ast)\;d\mathrm{z}=:\Bold{P}(\mathrm{z})\;d\mathrm{z}\;.
\end{equation}
$\mathcal{N}$ is a normalization constant, $\Bold{K}(\mathrm{z}',\mathrm{z}^\ast)\Bold{K}^{-1}(\mathrm{z},\mathrm{z}^\ast)=\delta(\mathrm{z}',\mathrm{z})\Bold{Id}$, and ${Id}$ is the identity operator on $\mathcal{H}$. From these, one obtains the local CS superposition on $\mathcal{W}$;
\begin{eqnarray}
\psi\rangle_ i
&:=&\int_{U_ i}|\mathrm{z}^\ast;\Bold{\mu})\; \mathbf{d\,\Bold{\sigma}}(\mathrm{z})\;(
{\mathrm{z}};\Bold{\mu}|\psi\rangle
\end{eqnarray}
which must then be extended globally to $\Z$. Similarly,
\begin{eqnarray}\label{operator symbol}
\langle\psi|{O}\,\psi\rangle_{\mathcal{H}}
&:=&\int_{\Z}{\psi}^\ast_{\Bold{\mu}'}(\mathrm{z}')\Bold{P}^\ast(\mathrm{z}')\,
\widehat{{O}}\,{\psi}_{\Bold{\mu}'}(\mathrm{z}')\;d\mathrm{z}'\notag\\
&=&\int_{\Z}\int_{\Z}{\psi}^\ast_{\Bold{\mu}'}(\mathrm{z}')\Bold{P}^\ast(\mathrm{z}')\,
(\mathrm{z}';\Bold{\mu}'|O|\mathrm{z}^\ast;\Bold{\mu})\,\Bold{P}(\mathrm{z}){\psi}_{\Bold{\mu}}(\mathrm{z})\;
d\mathrm{z}'\,d\mathrm{z}\,\notag\\
&=:&\int_{\Z}\int_{\Z}{\psi}^\ast_{\Bold{\mu}'}(\mathrm{z}')\,
(\Bold{K}_{O}(\mathrm{z}',\mathrm{z}^\ast))_{\Bold{\mu}'\,\Bold{\mu}}\,{\psi}_{\Bold{\mu}}(\mathrm{z})\;
d\mathrm{z}'\,d\mathrm{z}\;.
\end{eqnarray}
Observe that $(\Bold{K}(\mathrm{z}',\mathrm{z}^\ast))_{\Bold{\mu}'\,\Bold{\mu}}$ and $\Bold{P}(\mathrm{z})$ are non-trivial due to the non-trivial commutator $[\mathfrak{z}^+_{a},\mathfrak{z}^-_{b}]=-2(\eta_{ab}\mathfrak{h}+i\mathfrak{j}_{ab})$. (Appendix \ref{propagator} calculates an explicit expression.) With suitable restrictions, $\widehat{{O}}\,{\psi}_{\Bold{\mu}'}(\mathrm{z}')=(\mathrm{z}';\Bold{\mu}'|O\psi\rangle$ can be rendered a distribution, and  $\Bold{K}_{O}(\mathrm{z}',\mathrm{z}^\ast)$ can be interpreted as the CS model of the propagator on $\mathrm{Z}$, associated with operator ${O}$, that can be realized as a QM functional integral (recall example \ref{Klein-Gordon} and see appx. \ref{propagator}).

The problem is, coherent states furnish representations for $SO(5,\C)$ but the dynamical group is $G_D=\R^{1,3}\rtimes SL(2,\C)$. The challenge then is to uncover the effects of the group contraction on these CS and interpret them as physical particles. Consequences of group contraction on the UIRs of $SO(4,1)\rightarrow ISO(3,1)$ were rigorously studied by \cite{MN}. With their results for assurance, we will take a more heuristic approach in the next subsection.

\subsubsection{Physical interpretation}\label{physical interpretation}
Before contraction, the Whitney vector bundle can be combined to form two real vector bundles with $\mathrm{B}:=(\Z+\Z^\ast)$ and $\mathrm{P}:=(\Z-\Z^\ast)$  base spaces and typical fibers $\mathcal{W}^\mathrm{B}_{(\Bold{\mu})}=\mathcal{W}_{(\Bold{\mu})}+\mathcal{W}^\ast_{(\Bold{\mu})}$ and $\mathcal{W}^\mathrm{P}_{(\Bold{\mu})}
=i(\mathcal{W}_{(\Bold{\mu})}-\mathcal{W}^\ast_{(\Bold{\mu})})$.\footnote{$\Z$ is a complex manifold so it is equipped with a complex structure which determines $\Z^\ast$.}  Referring to the analysis of (\ref{second decomposition}), they can be interpreted as boost-type and momentum-type bundles. Accordingly, momentum-type physical states $[\psi^{(r)}]_{\mathrm{p}^+}\in L^2(\mathrm{P},\mathcal{W}^\mathrm{P})$, where $\mathrm{p}^+$ labels the action of $\mathfrak{z}^+$ on $\mathrm{P}$, can be represented by momentum-type CS ${\psi}_{{\Bold{{\mu}}}}({\mathrm{p}})
={\psi}_{\Bold{\mu}}(\mathrm{z}-\mathrm{z}^\ast)$ with $\mathrm{p}\in \mathrm{P}$; similarly for boost-type states. Before group contraction the two descriptions are symmetrical; not so post contraction.

From the adjoint representation of (\ref{first decomposition}), it is clear that $\mathfrak{z}^\pm$ are (oppositely) charged under $\mathfrak{h}$, and they constitute a symmetrical pair under complex conjugation. But group contraction breaks the symmetry between them, and the boost-type bundle becomes a less convenient description. If we suppose group contraction is somehow induced by dynamics, this has the flavor of dynamical symmetry breaking. Actually, it's more like dynamical symmetry \emph{deformation} in the sense that the group manifold structure is merely deformed.

It is helpful to have a model for the contraction: we will imagine the quantum number coming from the adjoint representation of $[\mathfrak{z}^\pm_{a},\mathfrak{h}]=\mp \mathfrak{z}^\pm_{a}$ corresponds to a unit wavenumber associated with $\mathfrak{z}^\pm_{a}$ on the oriented, additive circle group $SO(2)\cong S^1$ with radius $R$ generated by $\mathfrak{h}$. Hence, we interpret quantum numbers associated with $\mathfrak{h}$ as frequencies $\pm\omega$.\footnote{Of course $\mathfrak{h}$ is usually interpreted as the generator of dilatations.} Contraction by $\mathfrak{h}\rightarrow\mathfrak{h}/R$ sends $S^1\stackrel{R\rightarrow\infty}{\longrightarrow}\R$ which sends $\mathbb{Z}\ni \pm\omega\rightarrow p_0\in \R_{\geq0}\cup\R_{\leq0}\cong\R$. This serves to augment the CS parametrizing \emph{non-compact} base space $\mathrm{P}$, which is now generated by the four mutually commuting algebra elements $\mathfrak{p}_a,\mathfrak{h}$ where $\mathfrak{p}_a=(\mathfrak{z}^+_a-\mathfrak{z}^-_a)$. As $R\rightarrow\infty$, the expectation $\langle\psi|(\mathfrak{{z}}^++\mathfrak{{z}}^-)\,\psi\rangle$ vanishes regardless, but $\langle\psi|(\mathfrak{{z}}^+-\mathfrak{{z}}^-)\,\psi\rangle$ grows like $R$. We require then $\mathfrak{e}_{a4}\rightarrow \mathfrak{e}_{a4}/R$ to ensure $\langle\psi|[(\mathfrak{{z}}^+-\mathfrak{{z}}^-),\mathfrak{{h}}]\,\psi\rangle$ remains finite. Physical states are now $[\psi^{(j,\sigma)}]_{\mathrm{p}^+}\in L^2(\R\times i\R^3,\mathcal{W}^\mathrm{P})$, and from (\ref{first decomposition}) one sees that $\mathfrak{b}_a=(\mathfrak{z}^++\mathfrak{z}^-_a)$ generates boosts and $\mathfrak{p}_\mu:=\{\mathfrak{p}_a,\mathfrak{h}\}$ is an ideal with respect to boosts. After contraction then, the CS are ${\psi}_{{\Bold{\mu}}}(p)$
where ${\Bold{\mu}}\equiv (j,\sigma,\mathrm{p}^+)$ and $p\in\R^{1,3}$ with boost-invariant $p_0^2-\mathrm{p}^2=m_0^2$.  Thereby interpret the equivalence class $[\psi^{(j,\sigma)}]_{\mathrm{p}^+}$ as a collection of $\mathrm{dim}(\mathcal{W}^\mathrm{P}_{{\mu}})=2j+1$ states labeled by $\mathrm{p}^+$ and characterized by boost-invariant rest mass $m^2_0$ and spin $j$. But we can't say this represents a single particle --- there is no number operator here.

The case of $m_0^2=0$ has an intereseting wrinkle: Consider a ground state $\psi_0(\mathrm{z})\in\mathcal{W}_{(\Bold{\mu})}$.  The contracted triangular decomposition gives $(\rho'(\mathfrak{z}^+)\,\psi_0)(\mathrm{z})
=(\rho'(\mathfrak{e}_{0a}+i\mathfrak{e}_{a4}/R)\,\psi_0)(\mathrm{z})=0$  which implies $(\rho'(\mathfrak{e}_{0a})\,\psi_0)(\mathrm{z})=(\rho'(\mathfrak{b}_{a})\,\psi_0)(\mathrm{z})=0$ when $R\rightarrow\infty$. Furthermore, defining in the usual way $\mathfrak{j}_\pm:=\mathfrak{j}_{13}\mp i\mathfrak{j}_{23}$ and $\mathfrak{j}_0:=\mathfrak{j}_{12}$ and using $i[\mathfrak{z}^\pm_{a},\mathfrak{j}_{bc}]=\pm\eta_{ab}\mathfrak{z}^\pm_{c}
\mp\eta_{ac}\mathfrak{z}^\pm_{b}$, we see that\footnote{Technically, this relation allows $(\rho'(\mathfrak{j}_{-})\,\psi_0)(\mathrm{z})\propto\psi_0(\mathrm{z})$, but then we would have $\rho'(\mathfrak{j}_{-})\propto\rho'(\mathfrak{j}_{0})$ which would violate the $\mathfrak{j}_{ab}$ commutation relations.} $(\rho'(\mathfrak{j}_{-})\,\psi_0)(\mathrm{z})=0$ as well as  $(\rho'(\mathfrak{j}_{+})\,\psi_0)(\mathrm{z})=0$ and $(\rho'(\mathfrak{j}_0)\,\psi_0)(\mathrm{z})=\sigma\psi_0(\mathrm{z})$ trivially.  Conclude that, after contraction, ground states of spin $j$ and $\pm |p_0|$ are annihilated by the boost operator, and in consequence they are characterized by a single helicity instead of the $2j+1$ helicity states for $m_0^2>0$. Evidently, one should identify the (boost-invariant) ground states $\breve{\psi}_0\in\mathcal{H}$ with massless states, and the entire equivalence class $[\psi_0^{(j,\sigma)}]_{\mathbf{p}^+}$ represents a single massless state. With the exception of the vacuum state, these massless states represent degenerate helicity degrees of freedom post contraction. Notice the $m_0^2=0$ case analyzed here via the triangular decomposition (\ref{first decomposition}) is free of the continuous spin states that one encounters using the little group analysis of decomposition (\ref{second decomposition}).

Rehashing the method of {remark \ref{classical phase space}}, interpret $\mathbb{D}^{(\mu)}_p(t):=\left(\bar{\rho}'(\mathfrak{p}^\mu(t))\,\psi_0\right)(p)$ as a dynamical-time-dependent tangent space at $p$ of a momentum-configuration manifold $\mathbb{P}(t)$ with Lie bracket $[\mathfrak{p}_\mu(t),\mathfrak{p}_\nu(t)]=0$. Similarly, construct a boost-configuration manifold $\mathbb{B}(t)$ with Lie bracket $[\mathfrak{b}_a(t),\mathfrak{b}_b(t)]=0$. If the Hamiltonian has the dynamical-time-independent form $\mathrm{H}=1/2\mathfrak{p}_\mu\mathfrak{p}^\mu+\mathrm{V}(\mathfrak{b}_a,\mathfrak{p}_\mu)$, then $\langle\psi_0|\rho'(d\mathfrak{p}_\mu/dt)\psi_0\rangle
\sim\langle\psi_0|\rho'([\mathfrak{p}_\mu,\mathrm{V}(\mathfrak{b}_a,\mathfrak{p}_\mu)]
\,\psi_0\rangle$. Evidently, $\mathbb{P}(t)$ is static if $[\mathfrak{p}_\mu,\mathrm{V}(\mathfrak{b}_a,\mathfrak{p}_\mu)]\sim f(\mathfrak{z}^+_a)$, and we can form the product manifold $\mathbb{P} \times\mathbb{B}(t)$. The evolution of a \emph{free} state $[\psi^{(j,\sigma)}]_{\mathrm{p}^+}$ (i.e. $\mathrm{V}(\mathfrak{b}_a,\mathfrak{p}_\mu)=0$), propagated according to (\ref{explicit overlap}), with fixed initial and final momentum would then be associated with all continuous paths in $\mathbb{P}\times \mathbb{B}$ with the prescribed boundary conditions. If we can identify the equivalence-class representative $[\psi^{(j,\sigma)}]_{\mathrm{p}^+=0}$ as a \emph{particle}, then in general the particle content will change along a given path in $\mathbb{P}\times \mathbb{B}$ since $\Bold{K}(\mathrm{z}',\mathrm{z}^\ast)$ is non-trivial.

Alternatively, going back to {(\ref{second decomposition})} after contraction, one can form two Whitney vector bundles with base space $X:=(Z+Z^\ast)$ and $P:=(Z-Z^\ast)$. This case parallels the Heisenberg case. In particular, for an evolution Hamiltonian $\mathrm{H}\equiv1/2\mathfrak{p}_\mu\mathfrak{p}^\mu$ it yields a static cotangent bundle $T^\ast\mathbb{X}=\mathbb{{X}}\times\mathbb{P}$ and associated CS $\psi_{\Bold{\widetilde{\mu}}}(x)$ where now $\Bold{\widetilde{\mu}}\equiv(j,\sigma,\mathrm{x}^+)$ and $x\in\R^{1,3}$. How should one interpret $[\psi^{(j,\sigma)}]_{\mathrm{x}^+}\in L^2(\R^{1,3},\mathcal{W}^X)$ and its associated CS ${\psi}_{{\Bold{\widetilde{\mu}}}}(x)$?

One suggestion, inspired by QFT, is to view position-type CS as the mass-shell Fourier transform of a linear superposition of momentum-space CS:
\begin{equation}\label{position wave function}
 {\psi}_{\Bold{\widetilde{\mu}}}(x)
 :=\sum_{\sigma}\int_{\R^{1,3}}
 \left[u_{\Bold{\widetilde{\mu}}}(p,\sigma){\psi}_{\Bold{{\mu}}}^\ast(p)e^{ip.x}
 +v_{\Bold{\widetilde{\mu}}}(p,\sigma){\psi}_{\Bold{{\mu}}}(p)e^{-ip.x}\right]
 \theta(p_0)\delta(p^2-m_0^2)\,dp
\end{equation}
where ${\Bold{\widetilde{\mu}}}$ is a collection of suitable labels and $u_{\Bold{\widetilde{\mu}}}(p,\sigma)$ and $v_{\Bold{\widetilde{\mu}}}(p,\sigma)$ are sufficiently smooth functions whose Poincar\'{e} transformation properties are determined by those of ${\psi}_{\Bold{{\mu}}}(p)$ and the requirement that ${\psi}_{\Bold{\widetilde{\mu}}}(x)$ be Poincar\'{e} covariant with spin content $(j,\sigma)$. The propagator for ${\psi}_{\Bold{\widetilde{\mu}}}(x)$ then comes from the mass-shell Fourier transform of $\Bold{K}_{U_\tau}(p',p)$.

At this point, ${\psi}_{\Bold{\widetilde{\mu}}}(x)$ is just a CS. To get something resembling a quantum field, restrict to a single representation and identify $[\psi^{(j,\sigma)}]_{\mathbf{z}^+=0}$ with an elementary particle type; in which case ${\widetilde{\mu}}={\mu}=(j,\sigma)$. Then introduce particle/anti-particle observables $\mathrm{a}^+,\mathrm{a}^-\in UM(\mathbf{F}_{\mathbb{S}}(G^\C))$ and use {(\ref{pi_x})} to define associated CS \emph{creation} operators\footnote{We now have access to a particle number operator. Recall we posit dynamical invariance under the inner automorphism $\mathfrak{G}_+\leftrightarrow\mathfrak{G}_-$. Consequently, interpret the action of $a^-(p,\mu)=(a^+(p,\mu))^\dag=(a^+(p,\mu))^{-1}$ on the ground state as generating a CS associated with an anti-particle as opposed to annihilating the ground state. Note that $\mathrm{a}^-\mathrm{a}^+-\mathrm{a}^+\mathrm{a}^-\notin UM(\mathbf{F}_{\mathbb{S}}(G^\C))$ since it does not possess an inverse, and equation (\ref{group comutator}) holds even if $\mathfrak{G}_+ \nleftrightarrow\mathfrak{G}_-$. So they don't carry any Lie bracket structure of the underlying $C^\ast$-algebra $\mathfrak{A}_L$.}
\begin{eqnarray}
a^+(p,\mu){\psi}_0&:=&\pi_p(\mathit{\Pi}_\lambda^{(1)}(\mathrm{a}^+))\psi_0(p)
=((\mathit{\Pi}_\lambda^{(1)}(\mathrm{a}^+))\psi_0)(p)=\psi_{\mu}(p)\notag\\
a^-(p,\mu){\psi}_0&:=&\pi_p(\mathit{\Pi}_\lambda^{(1)}(\mathrm{a}^-))\psi_0(p)
=((\mathit{\Pi}_\lambda^{(1)}(\mathrm{a}^-))\psi_0)(p)=\psi^\ast_{\mu}(p)\;.\notag\\
\end{eqnarray}
Applying these to {(\ref{position wave function})} yields `CS fields' ${\Psi}_{\Bold{{\mu}}}(x)$ defined by ${\Psi}_{\Bold{{\mu}}}(x){\psi}_0:={\psi}_{\Bold{{\mu}}}(x)$, but these certainly are \emph{not} the quantum fields of a QFT --- although they do not commute.

Here we come to an important point: $\mathrm{a}^+,\mathrm{a}^-\in UM(\mathbf{F}_{\mathbb{S}}(G^\C))$ implies the \emph{group} commutators of the associated operators satisfy
\begin{equation}\label{group comutator}
[a^-(p,\mu),a^+(p,\mu)]_{G^\C}=Id\;,
\end{equation}
which happens to agree with commutation relations for creation/annihilation operators. But {(\ref{group comutator})} follows trivially from the fact that $\mathrm{a}^+,\mathrm{a}^-$ are unitary observables, and there is no reason to expect $[\rho'(\log \mathrm{a}^-),\rho'(\log \mathrm{a}^+)]_\pm={id}$ for their generators. Still, being representations of elements in $\mathbf{F}_{\mathbb{S}}(G^\C)$, the operators $a^+(p,\mu)$ and $a^-(p,\mu)$ transform under inner automorphisms of Poincar\'{e} in agreement with QFT, but they do not define the quantum algebra in contrast to QFT. This leads to a different interpretation of $a^+(p,\mu),a^-(p,\mu)$ (particle/anti-particle creation v.s. particle creation/annihilation), a different formulation of the quantum Hamiltonian, and a different picture of the quantum vacuum.

\subsubsection{Dynamics}
Given the CS model of physical states ${\psi}_{\Bold{\mu}}(p)$, consider system dynamics governed by some Hamiltonian $\mathrm{H}(t)=\mathfrak{C}\,t+\mathrm{V}(t)$ coming from the universal enveloping algebra $\mathrm{U}(\mathfrak{iso}(1,3))$ where $\mathfrak{C}$ is a linear combination of the two Casimirs $\mathfrak{C}_{P^2}$ and $\mathfrak{C}_{W^2}$ of the Poincar\'{e} group. For the evolution operator $U_\tau=\rho(e^{-(\mathfrak{C}\,\tau+\mathrm{V}(\tau)})$, we end up with transition amplitudes
 \begin{equation}\label{transitions}
\langle\phi|U_\tau\,\psi\rangle
=\int_{\R^{1,3}}\int_{\R^{1,3}}\Bold{\phi}^\ast_{\Bold{{\mu}}'}(p')\,
\Bold{K}_{U_\tau}(p',p)\,{\psi}_{\Bold{{\mu}}}(p)
dp\,dp'
\end{equation}
If $\mathrm{H}(t)$ commutes with $\mathfrak{iso}(1,3)$ and we identify evolution-time $\tau$ with proper time (or we agree not to associate  $\tau$ as a conjugate variable to the $\mathfrak{h}$ direction in $\R^{1,3}$), there is no tension with Poincar\'{e} invariance.\footnote{If we do insist (as usual) on associating $\tau$ with the $\mathfrak{h}$ coordinate, Poincar\'{e} invariance can still be maintained if the evolution operator satisfies $(p',\Bold{{\mu}}'|U_\tau|p,\Bold{{\mu}})\sim e^{ip\cdot q}(p',\Bold{{\mu}}'|\widetilde{U}|p,\Bold{{\mu}})$ --- which is generally the case in Poincar\'{e} QFT.}

The momentum space propagator $\Bold{K}_{U_\tau}(p',p)$ can be given a functional integral realization. Generically however, the propagator for the full Hamiltonian is difficult if not impossible to calculate. On the other hand, the interaction-free propagator $\Bold{K}_{e^{-\mathfrak{C}\,\tau}}(p',p)$, which is quadratic with commuting $\mathfrak{C}_{P^2}$, and $\mathfrak{C}_{W^2}$, \textit{can} be calculated explicitly. (Recall the elementary calculation of Klein-Gordon in example \ref{Klein-Gordon}.) Then Wick's theorem, together with $\Bold{K}_{e^{-\mathfrak{C}\,\tau}}(p',p)$ applied to the ground state, serves to construct a momentum-space perturbative  realization of $\Bold{K}_{U_\tau}(p',p)$ using standard methods. This provides a formulation of relativistic QM quite similar in spirit to \cite{Men}. But the dynamics generated by Poincar\'{e} can only alter $4$-momentum and helicity. To get a realistic theory with a Poincar\'{e}-invariant $S$-matrix describing Standard Model interactions, one clearly has to enlarge the dynamical group beyond Poincar\'{e}.

\subsection{Symplectic QM}
It is instructive to re-run the previous subsection with $G^\C=Sp(4,\C)$ which is the Langlands dual of $SO(5,\C)$. One might expect to find dual descriptions of recognizable objects.

The algebra $\mathfrak{sp}(4)$ is defined by the commutation relations
\begin{eqnarray}\label{commutation relations}
&&[\mathfrak{e}_{ab}\,,\,{\mathfrak{e}}_{cd}^\dag]
=\delta_{ac}\mathfrak{u}_{db}+\delta_{ad}\mathfrak{u}_{cb}
+\delta_{bc}\mathfrak{u}_{da}+\delta_{bd}\mathfrak{u}_{ca}\notag\\
&&[\mathfrak{u}_{ab}\,,\,{\mathfrak{u}}_{cd}]
=\delta_{bc}\mathfrak{u}_{ad}-\delta_{ad}\mathfrak{u}_{cb}\notag\\
&&[\mathfrak{u}_{ab}\,,\,{\mathfrak{e}}_{cd}]
=\delta_{bc}\mathfrak{e}_{ad}+\delta_{bd}\mathfrak{e}_{ac}\notag\\
&&[\mathfrak{u}_{ab}\,,\,{\mathfrak{e}}_{cd}^\dag]
=-\delta_{ac}{\mathfrak{e}}_{bd}^\dag
-\delta_{ad}{\mathfrak{e}}_{bc}^\dag\notag\\
&&[\mathfrak{e}_{ab}\,,\,{\mathfrak{e}}_{cd}]=
[{\mathfrak{e}}_{ab}^\dag\,,\,{\mathfrak{e}}_{cd}^\dag]=0
\end{eqnarray}
where $\mathfrak{e}_{ab}=\mathfrak{e}_{ba}$ with $a,b\in\{1,2\}$. This algebra can be constructed from bosonic creation/annihilation operators.\cite{JQ} Notice that the subalgebra $\{\mathfrak{u}_{ab}\}$ generates $U(2)$.

The decomposition
\begin{eqnarray}
&&\{\mathfrak{u}_{ab}\}
=:\{\mathfrak{h}_a,\mathfrak{u}_{+},\mathfrak{u}_{-}\}
,\hspace{1.4in}a\in\{1,2\}\notag\\
&&\{{\mathfrak{e}}_{ab},{\mathfrak{e}}_{ab}^\dag\}
=:\left\{(\mathfrak{e}_{a},\mathfrak{e}_{a,b}),
(\mathfrak{e}^\dag_{a},\mathfrak{e}^\dag_{a,b}) \right\} ,\;\;\;\;\;a,b\Rightarrow 1\leq a< b\leq2\
\end{eqnarray}
allows to define
$\mathfrak{G}_0\cong\mathrm{span}_\R\{\mathfrak{h}_a\}$,
$\mathfrak{G}_+\cong\mathrm{span}_\R\{\mathfrak{u}_{+},\mathfrak{e}_a,\mathfrak{e}_{a,b}\}$, and
$\mathfrak{G}_-\cong\mathfrak{G}_+^\dag$ which yields a triangular decomposition. The maximal compact subalgebra is $\mathfrak{u}(2)$ so we have the parabolic subalgebra $\mathfrak{P}^\C=\mathrm{span}_\C\{\mathfrak{\mathfrak{e}}_{ab}^\dag\}\cup
\mathrm{span}_\C\{\mathfrak{\mathfrak{u}}_{ab}\}$ leading to a principal and associated bundle with base space $\Bold{Z}=Sp(4,\C)/P^\C\cong M_2^{sym}(\C)$ where $M_2^{sym}(\C)$ is the vector space of complex-valued $2\times2$ symmetric matrices. The Hilbert space contains equivalence classes $[\psi^{(\bar{j},\bar{\sigma},\bar{n})}]_{\Bold{z}^+}$ where $(\bar{j},{\sigma},\bar{n})$ labels both bosonic and fermionic UIRs of $\mathfrak{u}(2)$ and the label ${\Bold{z}^+}\in M_2^{sym}(\C)$ comes from the right action of $\mathfrak{e}_{ab}^\dag$. The associated ``matrix'' CS are ${\psi}_{\bar{\Bold{{\mu}}}}({\Bold{z}})$. Being Langlands dual, one might anticipate a correspondence between ${\psi}_{{\bar{\Bold{{\mu}}}}}(\Bold{z})$ and ${\psi}_{{{\Bold{{\mu}}}}}({z})$, but it is not obvious. However, $\mathfrak{sp}(4,\C)\cong\mathfrak{so}(5,\C)$ so the correspondence should be discernable.

To progress,  recall that complex symmetric $\Bold{z}$ implies $\Bold{z}\Bold{z}^\ast\in M_2^{H^{(+)}}(\C)$ is a positive-semidefinite $2\times2$ Hermitian matrix. So rather than repeating the strategy for $SO(5,\C)$ by considering momentum-type $(\Bold{Z}-\Bold{Z}^\dag)\cong M_2^{H}(\C)$ and so on, we will cut to the chase and form a position-type vector space $\Bold{X}:=\Bold{Z}\Bold{Z}^\ast\cong M_2^{H^{(+)}}(\C)$ and identify $\Bold{x}\in\Bold{X}$ as
\begin{equation}
\Bold{x}\equiv\left(
     \begin{array}{cc}
       x_0+x_3 & x_1+ix_2 \\
       x_1-ix_2 & x_0-x_3 \\
     \end{array}
   \right)
\end{equation}
with $x_0,x_1,x_2,x_3\in\R$.  Notice $\det(\Bold{x})=x_0^2-x_1^2-x_2^2-x_3^2$ which we will identify with a spacetime interval $s_0^2$.

Since $\mathfrak{u}(2)\cong\mathfrak{su}(2)\oplus\mathfrak{u}(1)$, it is reasonable (from Langlands duality) to contract $\mathfrak{u}(1)$ in parallel with the $\mathfrak{so}(3)\oplus\mathfrak{so}(2)$ case. Similarly, duality suggests to contract $\mathfrak{e}_{ab}$. Contraction of $U(1)\stackrel{R\rightarrow\infty}{\rightarrow}\R$ yields physical state vectors $[\psi^{(\bar{j},\bar{\sigma})}]_{{\Bold{z}^+}}$ with a CS representation ${\psi}_{{\bar{\Bold{{\mu}}}}}(\Bold{z})$ where now $\Bold{z}\in \R\times M_2^{sym}(\C)$.  Since $M_2^{sym}(\C)$ is already a vector space over $\R$, contraction has no essential effect on $\Bold{X}$; although it renders $\Bold{Z}$ chargeless under $U(1)$ and alters the $[\mathfrak{e}_{ab}\,,\,{\mathfrak{e}}_{cd}^\dag]$ commutators. We suppose that post contraction we have $X_D=Sp(4,\R)/SL(2,\C)$. From $P_D=SL(2,\C)$ it follows that $\det(\Bold{x}'):=\det(S\Bold{x}S^\dag)=\det(\Bold{x})=s_0^2\geq0$ for all $S\in SL(2,\C)$, and recall that $G_D=Sp(4,\R)\cong Spin(3,2)$. This structure is clearly related to the spinor-helicity formalism (Fourier transformed), and we can identify the two eigenvectors of $\Bold{z}\in\Bold{Z}$ with spinor-helicity variables $\lambda^I_\alpha$ where $\alpha\in\{1,2\}$ and $I=1,2$ labels the eigenvector. The gulf between $s_0^2>0$ and $s_0^2=0$ is evident in the reduction of $\mathrm{rank}(\Bold{z})=2\rightarrow\mathrm{rank}(\Bold{z})=1$ when $\det(\Bold{x})=0$.

Apparently, in a loose sense $\Bold{Z}$ is the square root of Minkowski spacetime and ${\psi}_{{\bar{\Bold{{\mu}}}}}(\Bold{z})$ represents a set of $2j+1$ physical states labeled by $\Bold{z}^+$ and parametrized by spinor-helicity variables. We learn that points in the bundle $SL(2,\C)\rightarrow Sp(4,\C)\stackrel{\pi}{\rightarrow} \Bold{Z}$ efficiently encode scattering degrees of freedom, but the wave-function nature of  ${\psi}_{\bar{\Bold{{\mu}}}}({\Bold{z}})$ is obscured. Otherwise said,  $Sp(4,\C)$ underlies the spinor-helicity formalism.  This picture is dual to the bundle $SL(2,\C)\rightarrow SO(5,\C)\stackrel{\pi}{\rightarrow} \mathrm{Z}$ where the wave-function nature of  ${\psi}_{{\Bold{{\mu}}}}(\mathrm{z})$ is clear but scattering is obscured. Otherwise said,  $SO(5,\C)$ underlies the propagator formalism.

In this context, $Sp(4,\C)\stackrel{Langlands}{\longleftrightarrow} SO(5,\C)$ duality is not so much about strong/weak coupling: it is more about complementary descriptions of physical states ${\psi}_{\bar{\Bold{{\mu}}}}({\Bold{x}})\leftrightarrow{\psi}_{\widetilde{\Bold{{\mu}}}}({{x}})$ representing the dynamics induced by $Sp(4,\C)/SO(5,C)$. The symplectic model of scattering versus the orthogonal model of propagation echoes particle/wave duality. These two models can be combined under $OSp(5|4)$ to render a more democratic, hybrid spinor/vector parametrization of CS. But this SUSY, under the interpretation we espouse, does not imply a doubling of {physical} degrees of freedom. Rather, it doubles their mathematical manifestation as both particle \emph{and} wave as opposed to particle \emph{or} wave.\footnote{The thing is, it seems that an actual observation/measurement by an external agent can only ``see'' a particle \emph{or} a wave.}

\begin{remark}
It is known that $\mathfrak{u}(2)$ can also contract to the (bosonic and/or fermionic) oscillator algebra (see e.g. \emph{\cite{GIL}}))
\begin{eqnarray}
&&[N,a]=-a\notag\\
&&[N,a^\dag]= a^\dag\notag\\
&&[a,a^\dag]=-I
\end{eqnarray}
where $\mathfrak{h}_1\rightarrow N=a^\dag a$ and $\mathfrak{h}_2\rightarrow I$.

Whereas contraction produced an extra dimension in the CS parametrizing space for $SO(5,\C)$, in this case it  leads to a projective representation. This indicates $G_D$ is the metaplectic group $Mp(4,\R)$. After contraction and constructing real fiber bundles, physical state vectors are characterized by $[\psi^{(N,c)}]_{\mathbf{p}^+,\pm}$ where $\mathbf{p}\in\R^3$ and $c$ is a constant associated with the central extension. The $\pm$ corresponds to the two irreducible components of the metaplectic representation. In this sense, the parametrizing  space becomes $\R^3\times\mathbb{Z}_2$ after contraction, and ${\psi}_{{\Bold{\widetilde{\mu}}}}(\mathbf{p})$ represents two sets of harmonic oscillator degrees of freedom with quantum numbers $(N,c)$ located at the point $\mathbf{p}\in\R^3$. This facilitates a second quantization picture\footnote{It is remarkable that this contracted, non-relativistic picture supports mode creation and annihilation.} on $\R^3$, but we can't expect $[\psi^{(N,c)}]_{\mathbf{p}^+,\pm}$ to correspond directly to $[\psi^{(j,\sigma)}]_{\mathbf{b}}$ since the contractions differ.
\end{remark}

Return to $Sp(4,\C)$ before its contraction. Suppose one induces representations directly from $\mathfrak{u}(2)$. Then a real subspace of $\widetilde{Z}=Sp(4,\C)/U(2,\C)\cong M_2^{sym}(\C)\times M_2^{sym}(\C)$ would represent a \emph{non-commutative} phase space with $U(2)$ ``internal'' symmetry. Clearly $\widetilde{Z}$ won't parametrize a CS representation. But given a connection on the associated principal bundle, a $\mathrm{dim}_\C(3)$ integrable horizontal distribution could be constructed that would define a Lagrangian subspace $M_2^{sym}(\C)\in \widetilde{Z}$. The resulting structure would lead to a (matrix) QM theory of spinor-helicity degrees of freedom on $M_2^{sym}(\C)$ with CS characterized by ``external-like'' symmetries coming from the boost-type generators and ``internal-like'' symmetries coming from $U(2)$. This observation will be explored for $G^\C=Sp(8,\C)$ in a companion paper.

\section{Summary}
The assumptions enumerated in section \ref{preliminaries}, along with the functional integration framework summarized in appendix A, provide a hybrid realization of the axioms of quantum mechanics incorporating both functional and algebraic constructs. The center-piece of the construction is a topological group that:  (i) induces representations which directly determine the Hilbert space of states, (ii)  models the quantum $C^\ast$-algebra through the functional Mellin transform, (iii) suggests a topological interpretation of the measurement process, and (iv) generates dynamics through inner automorphisms on the $C^\ast$-algebra.

The topological group is the star, but the functional Mellin transform is the workhorse. Once the underlying topological group has been specified, functional Mellin simultaneously models the quantum $C^\ast$-algebra and provides representations of its observables, transition amplitudes, traces, and determinants. There is sophisticated mathematical machinery surrounding $C^\ast$-algebras and non-commutative function spaces, and the expectation is that functional Mellin will benefit from this and perhaps lead to useful computational techniques and methods in quantum physics beyond the simple examples presented here.

A notable aspect of the construction is the economy of assumptions that produce quantization. This is due to the leading role played by the topological group: it underlies both kinematics and dynamics. It means that one can replace the notion of `quantizing a classical system' with `specifying a topological group'. There is no ambiguity associated with the latter, and the role of the correspondence principle is reversed --- it now defines the `classical system' via the family $\Lambda$. But, there is no free lunch because one must still somehow determine the specific evolution observable and physical Hilbert space that yield the correct dynamics associated with a given quantum system.

We analyzed three well-known topological groups --- with varying degrees of detail. The construction applied to ordinary non-relativistic QM came out as expected and closely paralleled the standard treatment.

For relativistic QM we obtained Poincar\'{e}-invariant transition elements of invariant operators without the machinery of quantum fields. The break from QFT stems from the two different, underlying quantization methods. QFT starts with the {commutative} algebra of functions in $L^2(\R^{1,3},\mathcal{W}^\mathrm{P})$ then defines an operator--state correspondence via creation/annihilation operators acting on a Poincar\'{e}-invariant vacuum. This yields a {non-commutative} creation/annihilation operator algebra and, hence, a quantization. But the algebra does not contain the Poincar\'{e} group as units, and it does not carry the bracket structure of Poincar\'{e}. Nevertheless, by suitably restricting $u_{\Bold{\widetilde{\mu}}}(p,\sigma)$ and $v_{\Bold{\widetilde{\mu}}}(p,\sigma)$ in (\ref{position wave function}) and requiring Poincar\'{e} covariant $\psi_{\Bold{\widetilde{\mu}}}(x)$, QFT delivers a Poincar\'{e}-invariant $S$-matrix for physical particle scattering. Contrariwise, in our case $L^2(\R^{1,3},\mathcal{W}^\mathrm{P})$ serves only to carry the unitary representations used to construct the non-commutative $C^\ast$-algebra $\mathrm{F}_{\mathbb{S}}(G^\C)$ directly via functional Mellin. The output is a preferred Heisenberg picture of simple QM with position/momentum space serving only an interpretational role through a CS model allowing an explicit realization of expectation values. To paraphrase this important point: The Hilbert space of Poincar\'{e} IRs drives the construction of QFT. Quantization is achieved by establishing an {associated} Fock space acted on by creation/annihilation operators that transform suitably under Poincar\'{e}. This is the crucial step implementing $\mathrm{classical}\rightarrow\mathrm{quantum}$. In our approach, the $C^\ast$-algebra $\mathrm{F}_{\mathbb{S}}(G^\C)$ of Mellin integrable functions is determined by the group and the representation-furnishing Hilbert space. States can be reached by operating on the {ground state} in a given induced UR by the appropriate operators represented by functional Mellin transforms. In short, one could say that functional Mellin drives our construction, and the URs comprising the Hilbert space are a means to reveal various aspects of $\mathrm{F}_{\mathbb{S}}(G^\C)$. More succinctly, functional Mellin implements $\mathrm{quantum}\rightarrow\mathrm{classical}$.

Lastly, we looked at $Sp(4,\C)$ and found $Sp(4,\C)\stackrel{Langlands}{\longleftrightarrow} SO(5,\C)$ implies two complementary but equivalent descriptions of evolution dynamics: $Sp(4,\C)$ describes scattering while  $SO(5,\C)$ models correlations. Interestingly, the parametrizing base space of $SL(2,\C)\rightarrow Sp(4,\C)\stackrel{\pi}{\rightarrow} \Bold{Z}$, i.e. the spectrum of $\rho'(\mathfrak{e}_{ab})$, supplies a spinor-helicity model of spacetime; which therefore becomes a derived quantity. Also, spacetime only showed itself in the $SO(5,\C)$ description after Fourier transform of a CS model of physical states possessing spin and mass attributes. In both cases, spacetime becomes a secondary notion.

What has this tack gained us?\footnote{After all, the Poincar\'{e} group (and hence $SO(3,2)$) can't account for the Standard Model interactions. And even though functional Mellin can stand in for QFT for Poincar\'{e} QM (see appx. \ref{propagator} and \cite{Men,St,Sch}), it seems at this point that a QFT with a group restricted by the Coleman-Mandula theorem is required for any realistic theory, but we want to challenge that notion.} We outlined the functional Mellin model of the quantum mechanics of $SO(5,\C)$ and explored its contraction to the Poincar\'{e} group via $SO(3,2)$. In turn, it is known that Poincar\'{e} can be further contracted to the Galilean group. Inspection of (\ref{first decomposition}) shows that the real and imaginary parts of $\mathfrak{z}_a^\pm$  correspond to Galilean boosts and translations and $\mathfrak{j}_{ab}$ generate rotations. Consequently, for a holomorphic  scalar CS in $\C^3$, (\ref{first decomposition}) looks very much like the Heisenberg algebra augmented with the generators of rotation in $\R^3$. Indeed, the Heisenberg algebra (\ref{Heisenberg}) is a subalgebra of (\ref{first decomposition}) after a contraction effected by $R,\,S\rightarrow\infty$ with $\mathfrak{z}_a^\pm\rightarrow R^{-1}\,\mathfrak{z}_a^\pm$ and $\mathfrak{h}\rightarrow S^{-1}\,\mathfrak{h}$ where $S/R^2=\mathrm{constant}$. Furthermore, we have noted that $\mathfrak{sp}(4,\C)\cong\mathfrak{so}(5,\C)$ and $\mathfrak{sp}(4,\C)\supset\mathfrak{u}(2)$ contracts to the oscillator algebra. From this standpoint, the groups Poincar\'{e}, Galilean (with central charge), Heisenberg, and Oscillator (with central charge) represent different dynamical subgroups $G_D$ for an observer of a quantum system instantiating $G^\C\cong SO(5,\C)$. Then, according to our posit, all dynamics of the associated quantum mechanics is contained in $\mathit{\Pi}_\lambda^{(1)}(\mathrm{F}_{\mathbb{S}}(SO(5,\C)))$ with the various groups (Poincar\'{e}, Galilean, Heisenberg, and Oscillator) representing different stages of symmetry reduction (via subduction) and symmetry deformation (via contractions) determined by the particular system Hamiltonian.

Is there a group that encompasses the aforementioned --- including the Standard Model $SU(3)\times U(1)$ --- under contractions and subductions? We can think of two arguments that suggest the group is $Sp(8,\C)$: First, consulting the chain of subductions of a general $U(N)$ symmetry of a system having a very large number of $N$ constituents and then comparing to the known properties of elementary particles suggests the group is either symplectic or orthogonal, and phase space dynamics is symplectic. Second, the automorphism group of the dynamics of any realistic system is a symplectic group, and our dynamical Assumption 2.3 requires inner automorphisms; again singling out symplectic. The minimal group (with enough dimensions/degrees of freedom) that appears to suffice is $Sp(8,\C)$. From the lessons learned here, we expect its Langlands dual $SO(9,\C)$ is also relevant. Note that $\mathfrak{sp}(4,\C)$, $\mathfrak{so}(5,\C)$, and $\mathfrak{u}(4)\supset\mathfrak{su}(3)\times\mathfrak{u}(1)$ are all contained in $\mathfrak{sp}(8,\C)$, and one can anticipate CS based on something like $X_D=G_D/P_D=Sp(8,\R)/(Sp(4,\R)\times U(4))$ followed by suitable contractions might represent realistic Lorentzian physics. We will explore the quantum mechanics of $Sp(8,\C)$ via functional Mellin in a companion paper.

One final bit of speculation: We identify $SO(5,\C)$ ground states with massless states. This naturally leads to two copies (with frequency $\pm\omega$) of single-helicity massless particles with spin $j$. The vacuum, defined as the ground sate of the trivial representation of $SL(2,\C)$, is correspondingly immutable. Nevertheless, it can have a non-zero norm and is not annihilated by the `energy' observable $\mathfrak{h}$. How should we interpret the ground states on the $Sp(4,\C)$ side? We propose to identify the set of all ground states with the notion of ``vacuum'' in QFT. With the exception of the trivial representation, the action of $SL(2,\C)$ would create and/or leave unaltered particle/anti-particle constituents of scattering processes. On the other hand, the ground state that furnishes the trivial representation would remain unchanged; a spin-$0$ eigenstate of the one-dimensional Cartan subalgebra (post contraction) with a fixed norm $|v_{w^+}|$ and parametrized by spinor-helicity variables. If $Sp(4,\C)\stackrel{Langlands}{\longleftrightarrow} SO(5,\C)$ where the fundamental, kinematical starting point, one might expect to have $|v_{w^+}|=0$. Of course, as Poincar\'{e} doesn't include the strong and electroweak interactions this can't be the whole story. If instead the duality is reached, say, from the larger group duality $Sp(8,\C)\stackrel{Langlands}{\longleftrightarrow} SO(9,\C)$ through contraction and subduction by system dynamics, it is reasonable to expect $|v_{w^+}|\neq0$ generically. It is tempting to posit the dynamical $Sp(4,\C)\stackrel{Langlands}{\longleftrightarrow} SO(5,\C)$ vacuum state norm $|v_{w^+}|$ represents the cosmological constant. If so, $\Lambda$ won't suffer any quantum contributions from Poincar\'{e}.

\appendix
\section{Functional integration}\label{appx. A}
This appendix is a very brief summary of \cite{LA3}.

We are given the data $\left(G,\mathfrak{B},G_\Lambda\right)$
where $G$ is a Hausdorff topological group, $\mathfrak{B}$ is a
Banach space that may have additional algebraic structure, and
$G_\Lambda:=\{G_{\lambda},\lambda\in\Lambda\}$ is a family of
locally compact topological groups indexed by homomorphisms $\lambda:G\rightarrow G_{\lambda}$. The rigorous $\mathfrak{B}$-valued integration theory associated
with $\{G_{\lambda},\lambda\in\Lambda\}$ is used to define and characterize
functional integration on $G$.

\begin{definition}\label{int-def}
Let $\overline{\mathbf{F}}(G)$ represent a space of functionals  $\mathrm{F}:G\rightarrow\mathfrak{B}$, and denote the restriction of $\,\mathrm{F}$ to $G_\lambda$ by $f:=\mathrm{F}|_{G_\lambda}$. Let $\nu$ be a left Haar measure on $G_\lambda$.

A family (indexed by $\Lambda$) of integral operators
$\mathrm{int}_\Lambda:\overline{\mathbf{F}}(G)\rightarrow \mathfrak{B}$ is
defined by
\begin{equation}\label{FI}
\mathrm{int}_\lambda(\mathrm{F})=\int_G\mathrm{F}(g)\mathcal{D}_\lambda
g:=\int_{G_\lambda}f(g_\lambda)\;d\nu(g_\lambda)
\end{equation}
 given $f\in L^1(G_\lambda,\mathfrak{B})$ for all
$\lambda\in\Lambda$. We say that $\mathrm{F}$ is integrable with respect to the integrator family
$\mathcal{D}_\Lambda g$, and $\mathbf{F}(G)\subseteq\overline{\mathbf{F}}(G)$ is a space of integrable functionals.

Further, if $\mathfrak{B}$ is an algebra, define the functional $\ast$-convolution by
\begin{equation}
\left(\mathrm{F}_1\ast \mathrm{F}_2\right)_{{\lambda}}(g)
:=\int_G\mathrm{F}_1(\tilde{g})\mathrm{F}_2(\tilde{g}^{-1}g)
\mathcal{D}_{{\lambda}}\tilde{g}
\end{equation}
for each $\lambda\in\Lambda$.
\end{definition}

For any given
$\lambda$, the integral operator is linear and bounded according to
\begin{equation}
\|\mathrm{int}_\lambda(\mathrm{F})\|\leq\int_{G_\lambda}\|f(g_\lambda)\|
\;d\nu(g_\lambda)=\|f\|_{1,\lambda}<\infty\;.
\end{equation}
This suggests to define the norm $\|\mathrm{F}\|_{\mathbf{F}}:=\mathrm{sup}_\lambda\|\mathrm{int}_\lambda\mathrm{F}\|$.
The $\ast$-convolution then implies
\begin{eqnarray}\label{product}
\mathrm{int}_\lambda(\mathrm{F}_1\ast \mathrm{F}_2)
&=&\int_G(\mathrm{F}_1\ast \mathrm{F}_2)(g)\mathcal{D}_\lambda g\notag\\
&=&\int_{G_\lambda}\int_{G_\lambda}f_1(\tilde{g}_\lambda)
f_2(\tilde{g}_\lambda^{-1}g_\lambda)
\;d\nu(\tilde{g}_\lambda,g_\lambda)\notag\\
&=&\int_{G_\lambda}\int_{G_\lambda}f_1(\tilde{g}_\lambda)
f_2(g_\lambda)
\;d\nu(\tilde{g}_\lambda)d\nu(g_\lambda)\notag\\
&=&\int_{G_\lambda}\int_{G_\lambda}f_1(\tilde{g}_\lambda)
f_2(g_\lambda)
\;d\nu(\tilde{g}_\lambda)d\nu(g_\lambda)\notag\\
&=&\mathrm{int}_\lambda(\mathrm{F}_1)\,\mathrm{int}_\lambda(\mathrm{F}_2)
\end{eqnarray}
where the second line follows from left-invariance of the Haar measure and the last line follows from Fubini. A similar computation (using left-invariance of Haar and Fubini) establishes associativity $(\mathrm{F}_1\ast \mathrm{F}_2)\ast \mathrm{F}_3=\mathrm{F}_1\ast (\mathrm{F}_2\ast \mathrm{F}_3)$. Finally, given that $\mathfrak{B}$ is Banach, equation (\ref{product}) implies $\|\mathrm{F}_1\ast \mathrm{F}_2\|_{\mathbf{F}}\leq \|\mathrm{F}_1\|_{\mathbf{F}}\,\|\mathrm{F}_2\|_{\mathbf{F}}$. Consequently, $\mathbf{F}(G)$ inherits the algebraic structure of $\mathfrak{B}$:

\begin{proposition}\emph{(\cite{LA3}, Prop. 2.5)}
If $\mathfrak{B}\equiv\mathfrak{B}^\ast$ is a Banach $\ast$-algebra, then  $\mathbf{F}(G)$ --- when endowed with a suitable topology, an
involution $\mathrm{F}^\ast (g):={\mathrm{F}(g^{-1})}^\ast\Delta(g^{-1})$, and completed with respect to
the norm $\|\cdot\|_{\mathbf{F}}$ ---  is a Banach
$\ast$-algebra. Furthermore, $\mathrm{int}_\lambda$ is a $\ast$-homomorphism.
\end{proposition}

\begin{corollary}
If $\mathfrak{B}$ is a $C^\ast$-algebra, then $\mathbf{F}(G)$ is $C^\ast$-algebra.
\end{corollary}

Note that the products in $\mathbf{F}(G)$ and $L^1(G_\lambda,\mathfrak{B})$ are trivially equivalent
 by definition, but their respective norms are not. If the cardinality of $G_\Lambda$ is finite, our choice of norm on $\mathbf{F}(G)$  implies its restriction to $G_\Lambda$ is a direct sum
  $\mathbf{F}(G)|_{G_\Lambda}=\bigoplus_{\lambda\in\Lambda} L^1(G_\lambda,\mathfrak{B})$. In this regard, a `query' --- which
   corresponds to a particular $\lambda$ --- induces a projection.

\section{Poincar\'{e} reproducing kernel and Green's functions}\label{propagator}

\subsection{Reproducing kernel}
We want to find the normal-ordered form of the reproducing kernel of $SO(5,\C)$
\begin{eqnarray}\label{explicit overlap}
(\Bold{K}(\mathrm{z}',\mathrm{z}^\ast))_{\Bold{\mu}'\,\Bold{\mu}}
&:=&({\mathrm{z}'};\Bold{\mu}'|\mathrm{z}^\ast;\Bold{\mu})
=(\Bold{\mu}'|e^{
\sum_{a}{\mathrm{z}}'_{a}\,\mathfrak{z}^+_{a}}\,e^{
\sum_{b}{\mathrm{z}}^\ast_{b}\,\mathfrak{z}^-_{b}}|\Bold{\mu})\;.
\end{eqnarray}
First, write $\sum_{a}{\mathrm{z}}'_{a}\,\mathfrak{z}^+_{a}$ as $\frac{1}{2}\mathrm{tr}\,({\mathbf{z}'}\,\Bold{\mathfrak{z}}^+)$ where
\begin{equation}
\mathbf{z}'=\left(
              \begin{array}{cc}
                \mathrm{z}'_{11} & \mathrm{z}'_{12} \\
                \mathrm{z}'_{21} & \mathrm{z}'_{22} \\
              \end{array}
            \right)
=\left(
              \begin{array}{cc}
                2\,\mathrm{z}'_1 & \mathrm{z}'_2 \\
                \mathrm{z}'_2 & 2\,\mathrm{z}'_3 \\
              \end{array}
            \right)\;,\;\;\;\;\;\;
\Bold{\mathfrak{z}}^+
            = \left(
              \begin{array}{cc}
                \mathfrak{z}_{11}^+ & \mathfrak{z}_{12}^+ \\
               \mathfrak{z}_{21}^+ & \mathfrak{z}_{22}^+ \\
              \end{array}
            \right)
            = \left(
              \begin{array}{cc}
                \mathfrak{z}_1^+ & \mathfrak{z}_2^+ \\
               \mathfrak{z}_2^+ & \mathfrak{z}_3^+ \\
              \end{array}
            \right)\;.
\end{equation}

As this is a simple Lie group, we can proceed in any faithful representation. Because $\mathfrak{so}(5,\C)\cong\mathfrak{sp}(4,\C)$, it is convenient to work in the $4$-dimensional representation  where $[\Bold{\mathfrak{z}}^+]_{\alpha\beta}
=\mathrm{Id}_{2+\alpha,\beta}+\mathrm{Id}_{2+\beta,\alpha}$, and $[\Bold{\mathfrak{z}}^-]_{\alpha\beta}
=-[\Bold{\mathfrak{z}}^+]_{\beta\alpha}$, and $[\Bold{\mathfrak{s}}]_{\alpha\beta}
=\mathrm{Id}_{2+\beta,2+\alpha}-\mathrm{Id}_{\alpha,\beta}$ such that $\alpha,\beta\in\{1,2\}$, $\Bold{\mathfrak{s}}:=\{i\mathfrak{h},i\mathfrak{j}_{ab}\}$, and the $4\times4$ matrix $\mathrm{Id}_{i,j}$ has $1$ at position $(i,j)$ and $0$ everywhere else. For example, $\mathfrak{z}^+_1=[\Bold{\mathfrak{z}}^+]_{11}=2\mathrm{Id}_{3,1}$, and $\mathfrak{z}^+_2=[\Bold{\mathfrak{z}}^+]_{12}=\mathrm{Id}_{3,1}+\mathrm{Id}_{4,2}$, and $\mathfrak{z}^+_3=[\Bold{\mathfrak{z}}^+]_{22}=2\mathrm{Id}_{4,2}$. It is not hard to see that $\Bold{\mathfrak{z}}^+$ and $\Bold{\mathfrak{z}}^-$ are strictly triangular matrices and hence
\begin{equation}
e^{\frac{1}{2}\mathrm{tr}\,({\mathbf{z}'}\,\Bold{\mathfrak{z}}^+)}
=\left(
   \begin{array}{cc}
     \mathbf{1} & \mathbf{0} \\
     \mathbf{z}'& \mathbf{1} \\
   \end{array}
 \right)\;,\;\;\;\;\;\;\;\;
e^{\frac{1}{2}\mathrm{tr}\,({\mathbf{z}^\ast}\,\Bold{\mathfrak{z}}^-)}
=\left(
   \begin{array}{cc}
     \mathbf{1} & -\mathbf{z}^\ast \\
     \mathbf{0}& \mathbf{1} \\
   \end{array}\right)\;.
\end{equation}
Similarly, $\mathfrak{s}$ is $2\times2$ block diagonal so that
\begin{equation}
\mathrm{tr}\,({\mathbf{s}}\,\Bold{\mathfrak{s}})
=\left(
   \begin{array}{cc}
    \mathbf{s} & \mathbf{0} \\
     \mathbf{0} & -\mathbf{s}^{\mathrm{T}} \\
   \end{array}
 \right)
\end{equation}
and
\begin{equation}
e^{\mathrm{tr}\,({\mathbf{s}}\,\Bold{\mathfrak{s}})}
=\left(
   \begin{array}{cc}
     e^{\mathbf{s}} & \mathbf{0} \\
     \mathbf{0} & e^{-\mathbf{s}^{\mathrm{T}}} \\
   \end{array}
 \right)\;,\;\;\;\;\mathrm{with}\;\;\;\;\mathbf{s}:=\left(
                                          \begin{array}{cc}
                                            \mathrm{s}_{11} & \mathrm{s}_{13} \\
                                            \mathrm{s}_{23} & \mathrm{s}_{22} \\
                                          \end{array}
                                        \right)\;.
\end{equation}
Hence, from the BCH expansion we have
\begin{equation}
\Bold{K}(\mathrm{z}',\mathrm{z}^\ast)
=\left(
   \begin{array}{cc}
     \mathbf{1} & -\mathbf{z}^\ast \\
     \mathbf{z}' & \mathbf{1}-\mathbf{z}'\mathbf{z}^\ast \\
   \end{array}
 \right)
 =\left(e^{\frac{1}{2}\mathrm{tr}\,(\mathbf{r}\,\mathfrak{z}^-)}
 \;e^{\mathrm{tr}\,({\mathbf{s}}\,\Bold{\mathfrak{s}})}
 \;e^{\frac{1}{2}\mathrm{tr}\,(\mathbf{t}\,\mathfrak{z}^+)}\right)
\end{equation}
for some $2\times2$ matrices $\mathbf{r},\mathbf{s},\mathbf{t}$. Equating $2\times2$ blocks on each side yields
\begin{eqnarray}
\mathbf{1}&=&e^{\mathbf{s}}-\mathbf{r}\,e^{-\mathbf{s}^{\mathrm{T}}}\,\mathbf{t}\notag\\
\mathbf{z}^\ast&=&\mathbf{r}\,e^{-\mathbf{s}^{\mathrm{T}}}\notag\\
\mathbf{z}'&=&e^{-\mathbf{s}^{\mathrm{T}}}\,\mathbf{t}\notag\\
\mathbf{1}-\mathbf{z}'\mathbf{z}^\ast&=&e^{-\mathbf{s}^{\mathrm{T}}}\;.
\end{eqnarray}
Equivalently, $\mathbf{s}^{\mathrm{T}}=\log(\mathbf{1}-\mathbf{z}'\mathbf{z}^\ast)^{-1}$, $\mathbf{r}=\mathbf{z}(\mathbf{1}-\mathbf{z}'\mathbf{z}^\ast)^{-1}$, and $\mathbf{t}=(\mathbf{1}-\mathbf{z}'\mathbf{z}^\ast)^{-1}\mathbf{z}'$ as long as $\mathbf{z}'\mathbf{z}^\ast\in B^{open}_1(\mathbf{1})$.

Finally, for any representation $r$ with generators
$\{{\mathfrak{z}^+}^{(r)},{\mathfrak{z}^-}^{(r)},\mathfrak{s}^{(r)}\}$,
\begin{equation}
(\Bold{K}^{(r)}(\mathrm{z}',\mathrm{z}^\ast))_{\Bold{\mu}'\Bold{\mu}}
=(\Bold{\mu}'|\det(e^{{\mathbf{s}(\mathrm{z}',\mathrm{z}^\ast)}
\,\Bold{\mathfrak{s}}^{(r)}})|\Bold{\mu})
\end{equation}
where $\det$ applies to the $2\times2$ matrix and $\mathfrak{z}^+|\Bold{\mu})=0$ because $|\Bold{\mu})=\mathfrak{p}^\C|\Bold{v}_{w_+})$ and $[\mathfrak{z}^+,\mathfrak{p}^\C]\sim\mathfrak{z}^+$ for all $\mathfrak{p}^\C\in\mathfrak{P}^\C$. Under contraction, and again using the 4-dimensional representation with $\mathfrak{h}=i(\mathrm{Id}_{1,1}+\mathrm{Id}_{2,2}-\mathrm{Id}_{3,3}-\mathrm{Id}_{4,4})$ and putting $\lim_{R\rightarrow0} \frac{1}{2}(s_{11}(\mathrm{z}',\mathrm{z}^\ast)+s_{22}(\mathrm{z}',\mathrm{z}^\ast))/R=\beta$ for $\beta$ a constant, this reduces to
\begin{equation}
(\Bold{K}^{(r)}(\mathrm{z}',\mathrm{z}^\ast))_{\Bold{\mu}'\Bold{\mu}}
=(\Bold{\mu}'|e^{-i\beta\mathfrak{h}^{(r)}}\,e^{-i\sum_{a,b}{\mathrm{j}_{ab}(\mathrm{z}',\mathrm{z}^\ast)}
\,\Bold{\mathfrak{j}}_{ab}^{(r)}}|\Bold{\mu})\;,\;\;\;\;\;\;1\leq a< b\leq3
\end{equation}
where $\mathrm{j}_{12}(\mathrm{z}',\mathrm{z}^\ast)
=\frac{1}{2}(\mathrm{s}_{11}(\mathrm{z}',\mathrm{z}^\ast)-\mathrm{s}_{22}(\mathrm{z}',\mathrm{z}^\ast))$ and $\mathrm{j}_{a3}(\mathrm{z}',\mathrm{z}^\ast)=\mathrm{s}_{a3}(\mathrm{z}',\mathrm{z}^\ast)$.

\subsection{Green's functions}
Here we calculate the Green's functions for the Klein-Gordon operator of massive singlet coherent states in $L^2(\R^{1,3},\mathcal{W})$ induced from $P_D=SL(2,\C)$ for spins $0,1/2,1$. We follow \cite{PDM,LA6} which provides a path integral realization for elementary kernels of linear, second order partial differential equations on homogenous spaces $X=G/H$; which is relevant to our case $X_D=G_D/P_D$.(see \cite[appx. B]{LA6}) Intuitively, it calculates the Mellin transform  of the effective action of a QM point particle on $X$ in terms of an associated propagator on $G$.\footnote{In certain cases (specifically, unbounded manifold $X$) the path integral is quite similar to Schwinger's proper time method.} Of course the Green's functions can be alternatively realized as correlation functions via functional integrals over {fields} on $X$ according to QFT in the usual way, but it has long been realized \cite{St} that simple point-particle {path integrals} can also do the job---even with background gauge fields included. It goes by the name of worldline formalism these days.

The first step is to calculate the free scalar effective action. This is a simple exercise and it has been derived countless times using several different methods, but we include our version here for completeness and to emphasize the utility of {path integrals} for solving second order partial differential equations. Notation from and familiarity with \cite{PDM,LA6} is assumed.  The equation to solve is
\begin{equation}
\left(\frac{s^2}{4\pi}\square+V(x)\right)\Bold{K}_{_{\!\!KG}}(x_a,x_b)=-\Bold{\delta}(x_a-x_b)
\end{equation}
where $s\in\C_+$ is a constant, $V(x)=m^2$ in our case, $x_b,\;x_a\in\R^{1,3}$, and the point-to-point \emph{Dirichlet} elementary kernel is
\begin{eqnarray}
\Bold{K}_{_{\!\!KG}}(x_b,x_a)
&:=&\langle \psi_{x_b}|\Bold{K}_{_{\!\!KG}}\,\psi_{x_a}\rangle\notag\\
&=&\langle \psi_{x_b}|\mathcal{M}_\Gamma[\mathrm{E}^{-(\square+m^2)};1]\,\psi_{x_a}\rangle \notag\\ &=&\mathcal{M}_\Gamma[\langle \psi_{x_b}|\mathrm{E}^{-(\square+m^2)}\, \psi_{x_a}\rangle;1]\;.
\end{eqnarray}
In terms of CS, (\ref{operator symbol}) yields
\begin{eqnarray}
\langle \psi_{x_b}|\mathrm{E}^{-(\square+m^2)}\,\psi_{x_a}\rangle\notag\\
&&\hspace{-1.65in}=\int_{\R^{1,3}}\int_{\R^{1,3}}
\psi_{x_b}(z'_b)(z'_b;\Bold{\mu}'|
\mathrm{E}^{-(\square+m^2)}|z^\ast_a;\Bold{\mu})\psi_{x_a}(z^\ast_a)
\;dz'_b\,dz_a\notag\\
&&\hspace{-1.65in}=\int_{\R^{1,3}}\int_{\R^{1,3}}
\delta_\epsilon({x_b},\Re(z'_b))\Bold{K}_{_{\!\!KG}}(z'_b,z_a^\ast)
\delta_\epsilon({x_a},\Re(z_a))\;dz'_b\,dz_a\notag\\
\end{eqnarray}
with $z'_b:=z'(t_b)$ and $z^\ast_a:=z^\ast(t_a)$, and fixed-point coherent states $\psi_{x_b}(z'_b)=\delta_\epsilon({x_b},\Re(z'_b))$ and $\psi_{x_a}(z^\ast_a)=\delta_\epsilon({x_a},\Re(z_a))$. Here, $\delta_\epsilon$ denotes a \emph{regularized} delta distribution, and the chosen regularization must respect the covariance with respect to $P_D$. The limit $\epsilon\rightarrow0$ is only taken after integration, and generically one must introduce a renormalization scale $\mu$ relative to $\epsilon$ to render a finite quantity.

Further, for pointed maps $\mathcal{Z}\ni \zeta:(\mathbb{T},t_a)\rightarrow(\C^4,0)$ where $\mathbb{T}\equiv[t_a,t_b]\subset i\R$,
\begin{equation}\label{distribution}
\Bold{K}_{_{\!\!KG}}(z'_b,z^\ast_a)
:=\int_{\mathcal{Z}}\Bold{\delta}(z(t_b,\zeta)-z_b')e^{-\frac{1}{s}(\pi Q(z(t_b,\zeta))-\int_{t_a}^{t_b}m^2\;dt)}
\;\mathcal{D}\zeta
\end{equation}
with $z(t,\zeta)=z^\ast_a\cdot\mathit{\Sigma}(t-t_a,\zeta)$ where $\mathit{\Sigma}(t-t_a,\zeta)$ is a global transformation on $\R^{1,3}$ such that $z(t_a,\zeta)=z^\ast_a\cdot\mathit{\Sigma}(0,\zeta)=z^\ast_a$, and the Poincar\'{e} invariant and (gauge fixed) time-reparametrization invariant action is
\begin{equation}
 \pi Q(z(t_b,\zeta)-\int_{t_a}^{t_b}m^2\;dt
 =\int_{t_a}^{t_b}
z^\mu(t,\zeta)\left(\eta_{\mu\nu}\frac{d^2}{dt^2}\right)z^\nu(t,\zeta)\;dt
-\int_{t_a}^{t_b}m^2\;dt\;.
\end{equation}
Explicitly, the parametrization $z(t,\zeta)=z_{cr}(t)+s\zeta(t)$ is a variation about the critical path $z_{cr}(t)$ with fixed end-points  $z^\ast_a,z'_b$ given by $z_{cr}(t)=z^\ast_a+\left(\frac{z'_b-z^\ast_a}{t_b-t_a}\right)(t-t_a)$. Substituting into the functional integral yields the well-known heat kernel
\begin{eqnarray}
 \langle \psi_{x_b}|\mathrm{E}^{-(\square+m^2)}\psi_{x_a}\rangle
 &=&e^{-\frac{\Delta t}{s}\left(\frac{\pi(x_b-x_a)^2}{\Delta t^2}-m^2\right)}
 \int_{\mathcal{Z}}\delta(s\zeta(t_b))e^{-\frac{\pi}{s}\int_{t_a}^{t_b}
(s\dot{\zeta}(\tau))^2\;d\tau}\;\mathcal{D}\zeta\notag\\
&=&e^{-\frac{\Delta t}{s}\left(\frac{\pi(x_b-x_a)^2}{\Delta t^2}-m^2\right)}s^{-1}(s\Delta t)^{-2}
\end{eqnarray}
where $\Delta t:=t_b-t_a$ and we used
\begin{equation}
\delta(s\zeta(t_b))
=\int_{\R^{1,3}}e^{-2\pi i\langle su',\zeta(t_b)\rangle}\;du'
=\int_{\R^{1,3}}e^{-2\pi i\langle su'\delta_{t_b},\zeta\rangle}\;du'
=s^{-1}\int_{\R^{1,3}}e^{-2\pi i\langle \tilde{u}'\delta_{t_b},\zeta\rangle}\;d\tilde{u}'
\end{equation}
and skipped a few elementary steps that explicate $\delta_\epsilon$ and calculate the function integral.\footnote{The skipped steps include the localization $\lambda:\mathcal{Z}\rightarrow\R^{1,3}$ by $s\zeta\mapsto s\zeta(t_b)=:s{u}$ for ${u}\in\R^{1,3}$. By duality then, $\zeta'\mapsto s{u}'\delta_{t_b}$ which renders the covariance $\frac{-\pi}{s}W(s{u}'\delta_{t_b})=\frac{-\pi}{s}|s{u}'|^2\Delta t$.  Remark that, following localization, $\Bold{K}_{_{\!\!KG}}(z'_b,z^\ast_a)$ should be regarded as a distribution; in which case the action functional must localize to an honest function and this will typically involve regularization/gauge fixing and renormalization. Also, for the regularized delta distribution we used $\delta_\epsilon(x)=(s\epsilon)^{-1/2}e^{-\pi x^2/s\epsilon}$, and the limit $\epsilon\rightarrow0$ doesn't require renormalization in this example.} Similarly, by Fourier transform
\begin{eqnarray}
\langle \psi_{p_b}|\mathrm{E}^{-(\square+m^2)}\psi_{p_a}\rangle\notag\\
&&\hspace{-1.75in}=(\sqrt{2\pi})^4\delta(p_b-p_a)\frac{i(s\Delta t)^2}{4\pi^2}e^{-\frac{\Delta t}{4\pi s}(p_b^2s^2-4\pi m^2)}s^{-1}(s\Delta t)^{-2}\;.
\end{eqnarray}

For the next step, consider the operator
\begin{equation}
\mathcal{M}_\Gamma[\mathrm{E}^{-(\square+m^2)};\alpha]
=\int_{\phi_{\,\mathfrak{h}_{\mathrm{U}}}(\R)} e^{-(\square+m^2)(g)}\,\rho(g^\alpha)\;d\nu(g_{\Gamma})\;.
\end{equation}
It depends on $\rho$, and the one-parameter subgroup $\phi_{\mathfrak{h}_{\mathrm{U}}}(\R)$ characterizes evolution of a quantum system. If we want to talk about \emph{particle} transitions, we need $\rho$ to be a unitary \emph{irreducible} representation furnished by $\bigoplus_r\mathcal{V}_{(\mu)}^{(r)}$, and we need for it to act trivially on CS state vectors corresponding to elementary particles since the notion of `elementary' implies invariance under evolution. To characterize CS elementary particles, consider the two special evolution operators
\begin{equation}
\widehat{e^{-i H_{P}\tau}}\,{\psi}_{\Bold{\mu}}(z)
=({z};\Bold{\mu}|\rho(e^{-i\mathfrak{C}_{P^2}\tau})\,\psi\rangle\;,\;\;\;
\widehat{e^{-i H_{W}\tau}}\,{\psi}_{\Bold{\mu}}(z)
=({z};\Bold{\mu}|\rho(e^{-i\mathfrak{C}_{W^2}\tau})\,\psi\rangle
\end{equation}
where $\mathfrak{C}_{P^2}$ and $\mathfrak{C}_{W^2}$ are the two Casimirs associated with $\R^{1,3}\rtimes SL(2,\C)$. Since these evolutions leave ${\psi}_{\Bold{\mu}}(z)$ intact, it is reasonable to associate \emph{elementary} particles with eigenvectors obeying $\bar{\varrho}'(\mathfrak{C}_{P^2})\Bold{v}_{(m_0,j)}=m_0^2\Bold{v}_{(m_0,j)}$ and $\bar{\varrho}'(\mathfrak{C}_{W^2})\Bold{v}_{(m_0,j)}=-m_0^2j(j+1)\Bold{v}_{(m_0,j)}$ where $\bar{\varrho}$ is the inducing representation coming from $P^\C$ and $\Bold{v}_{(m_0,j)}\in\mathcal{V}_{(\mu)}^{(r)}$. The CS model of an elementary particle is then $\psi_{(m_0,j)}(z):=(z;\Bold{v}_{(m_0,j)}|\psi\rangle$.

Accordingly, we now take $\lambda:\phi_{\mathfrak{h}_{\mathrm{U}}}(\R)\rightarrow A_{\mathfrak{h}_{\mathrm{U}}}$ where $A_{\mathfrak{h}_{\mathrm{U}}}$ is abelian. This induces a UIR $g_\lambda\equiv a\rightarrow\rho(a)$ with $a\in A_{\mathfrak{h}_{\mathrm{U}}}$ such that $\widehat{\rho(a)}\psi_{(m_0,j)}(p)=\psi_{(m_0,j)}(p)$ for all $p\in\R^{1,3}$. In light of this, for each relevant representation $r$ let us take for the Hamiltonian operator
\begin{equation}
\mathrm{H}^{(r)}(a)=(p^2-m^2)\rho^{(r)}(a)=(p^2-m^2)\,\tau P_{\|}^{(r)}/|P_{\|}^{(r)}{P_{\|}^{(r)}}^\dag|
\end{equation}
where $\tau P_{\|}^{(r)}\in L(\mathcal{H})$, $\tau\in i\R_+\cup i\R_-$, and $P_{\|}^{(r)}$ is a projection onto an irreducible subspace of $\mathcal{H}$ furnished by $\bigoplus_r\mathcal{V}_{(\mu)}^{(r)}$. To verify the unitarity of representations $\rho^{(r)}$, notice that $\log (i|\tau|)\in i\R$ implies $\tau^\dag=\tau^{-1}=-i|\tau|^{-1}$ so that ${\rho^{(r)}(a)}^\dag={(\tau P_{\|}^{(r)}/|P_{\|}^{(r)}{P_{\|}^{(r)}}^\dag|)}^\dag
=-i\frac{|\tau|^{-1}}{|P_{\|}^{(r)}{P_{\|}^{(r)}}^\dag|}{P_{\|}^{(r)}}^\dag$ and therefore $\rho^{(r)}(a){\rho^{(r)}(a)}^\dag=Id$. (Recall from \S\ref{Hamiltonians} that $\tau^\dag\neq \tau^\ast$.)

The relevant objects to analyze are the total Hamiltonian
\begin{equation}
H(a):=\bigoplus_r \mathrm{H}^{(r)}(a)
=\bigoplus_r(p^2-m^2)\frac{\tau P_{\|}^{(r)}}{|P^{(r)}_{\|}{P^{(r)}_{\|}}^\dag|}
\end{equation}
and its elementary kernel
\begin{equation}
\left(\Bold{K}_{_{\!\!KG}}(p',p)\right)_{\Bold{\mu}'\Bold{\mu}}
=\bigoplus_r\left(\Bold{K}_{_{\!\!KG}}^{(r)}(p',p)\right)_{\mu'\mu}\;.
\end{equation}
We saw in \S\ref{Poincare} that $\R^{1,3}$ naturally parametrizes momentum-type CS. This was dictated by the Lie algebra structure of $\mathfrak{P}_D$. Consequently the interpretation of $\Bold{K}_{_{\!\!KG}}(p',p)$ is that it realizes point-to-point transitions on `momentum space' as a direct sum of propagators relative to the UIR furnished by $\mathcal{W}_{\Bold{\mu}}$.

The $SL(2,\C)$ singlet case was already treated in Example \ref{Klein-Gordon} where $\tau P_{\|}^{(0)}=\tau$ with $\tau\in i\R_+\cup i\R_-$. As a simple check we calculate
\begin{eqnarray}
\Bold{K}_{_{\!\!KG}}^{-\alpha}(p',p)
&=&\langle\psi_{p'}|\mathcal{M}_\Gamma[\Bold{K}_{_{\!\!KG}};\alpha]\psi_p\rangle\notag\\
&=&\frac{1}{\Gamma(\alpha)}\int_{i\R_+\cup i\R_-}\delta(p',p)e^{-\frac{i}{p^2-m^2}\tau}\tau^\alpha\;d\log (\tau)\notag\\
&=&e^{\frac{-i\pi\alpha}{2}}\left(\frac{1}{p^2-m^2}\right)^{-\alpha}\,
\delta(p',p)\;,\;\;\;\Re(\alpha)>0\;.
\end{eqnarray}
In particular, for $\alpha=1$, this yields $\Bold{K}_{_{\!\!KG}}^{-1}(p',p)=-i(p^2-m^2)\,\delta(p',p)$ which verifies the identity $\int\Bold{K}^{-1}(p',p'')\Bold{K}(p'',p)\;dp''=\delta(p',p)\Bold{Id}$.

\begin{remark}
We should point out that generically $O^{-\alpha}\neq(O^{-1})^\alpha$ (the present example notwithstanding). The latter can be represented as\emph{\cite{LA2}}
\begin{equation}
(O^{-1})_{H,\beta}^{\alpha}
:=\mathcal{M}_{H}
\left[(\mathrm{Id}-\mathrm{O})_{(\beta)};\alpha\right]
:=\int_{G^\C}\,\left(Id-\mathrm{O}(g)\right)^{-\beta}g^\alpha\;\mathcal{D}_{H} g\;,\;\;\;\;\;\;\;\alpha<\beta\in\mathbb{S}_{\lambda}\;.
\end{equation}
In the case at hand this gives
\begin{eqnarray}
(\Bold{K}_{_{\!\!KG}}^{-1})_{H,\beta}^\alpha
&=&
\int_{i\R_+\cup i\R_-}\delta(p',p)\left(1-\frac{i}{p^2-m^2}\tau\right)^{-\beta}\tau^\alpha\;d\log (\tau)\notag\\
&=&{\mathrm{B}(\alpha,\beta-\alpha)} e^{\frac{-\pi i\alpha}{2}}(p^2-m^2)^\alpha\, \delta(p',p)\;,\;\;\;\;\;\;0<\alpha<\Re(\beta)
\end{eqnarray}
for $\mathrm{B}(\cdot,\cdot)$ the beta function, and only one integral contributes depending on $\Re(p^2-m^2)\lessgtr 0$. In particular, for $\beta=2$ (recall $\alpha=1$) we get $(\Bold{K}_{_{\!\!KG}}^{-1})_{H,2}=-i(p^2-m^2)\,\delta(p',p)$ as expected.
\end{remark}

For higher spin state vectors labelled by $(j',j)$, the work to find the projection operators has already been done many times over and we just state the $(1/2,0)\oplus(0,1/2)$ and $(1/2,1/2)$ results.  For Dirac spinors we take
\begin{equation}
\frac{\tau P_{\|}^{(1/2)}}{|P_{\|}^{(1/2)}(P_{\|}^{(1/2)})^\dag|}=\frac{\tau}{p^2-m^2}(\slashed{p}\pm m)\;.
\end{equation}
which gives
\begin{eqnarray}
\left(\Bold{K}_{_{_{\!\!KG}}}^{(1/2)}(p',p)\right)_{\alpha\dot{\alpha}}
&=&\left(\slashed{p}\pm m\right)_{\alpha\dot{\alpha}}\,\langle \psi_{p'}|\mathrm{E}^{-\mathrm{H}^{(0)}}\psi_{p}\rangle\notag\\
&=&\frac{i\left(\slashed{p}\pm m\right)_{\alpha\dot{\alpha}}}{p^2-m^2}\,\delta(p'-p)\;.
\end{eqnarray}
For vector bosons, the operator $\square+m^2$ corresponds to the unitary gauge so we choose $P_{\|}^{(1)}=(\eta-(p\cdot p)/m^2)$ which yields
\begin{eqnarray}
\left(\Bold{K}_{_{\!\!KG}}^{(1)}(p',p)\right)_{\mu\nu}
&=&\left(\eta_{\mu\nu}-\frac{p_\mu p_\nu}{m^2}\right)\,\langle \psi_{p'}|\mathrm{E}^{-\mathrm{H}^{(0)}}\psi_{p}\rangle\notag\\
&=&\frac{i\left(\eta_{\mu\nu}-\frac{p_\mu p_\nu}{m^2}\right)}{p^2-m^2}\,\delta(p'-p)\;.
\end{eqnarray}
In both calculations we used the Haar measure to re-scale $\frac{\tau}{|P_{\|}^{(r)}(P_{\|}^{(r)})^\dag|}\rightarrow \tau$ and the operator identity
\begin{equation}
e^{-(p^2-m^2)\tau P_{\|}^{(r)}}\,\tau P_{\|}^{(r)}=e^{-(p^2-m^2)\tau}\,\tau P_{\|}^{(r)}
\end{equation}
which can be seen by expanding the exponential and using ${P_{\|}^{(r)}}^2=P_{\|}^{(r)}$.

Of course the examples we displayed in this appendix are well-known and standard: Our aim was to show the CS model of dynamical $\Bold{K}$ given by
\begin{eqnarray}
(z';\Bold{\mu}'|\Bold{K}_{U_\tau}|z^\ast;\Bold{\mu})
&:=&(z';\Bold{\mu}'|U_\tau^{-1}\,\Bold{K}\,U_\tau|z^\ast;\Bold{\mu})\notag\\
&=&(z'_\tau;\Bold{\mu}'|\,\Bold{K}\,|z^\ast_\tau;\Bold{\mu})\notag\\
&=:&(\Bold{K}(z'_b,z^\ast_a))_{\Bold{\mu}'\Bold{\mu}}
\end{eqnarray}
efficiently consolidates evolution of all relevant elementary particle species with respect to $G_D$ via functional Mellin. These methods hold more generally for particle/particle and particle/background field  interactions.\cite{St,Sch,Va}

\section{Complex time}\label{complex time}
Recall (from \S\ref{Hamiltonians}) the notion that one-parameter subgroups of  $G^\C$ correspond to system evolution. Here we want to briefly discuss the suggestion in footnote \ref{time inversion} to probe $G^\C$ with \emph{complex} one-parameter subgroups.

In section \ref{Hamiltonians}, an element  $g\in\phi_{\,\mathfrak{h}_{\mathrm{U}}}(\R)$ was characterized by localizing onto $\R_+\times \{i,-i\}$. We argued that $\tau Id\equiv\rho(g)$ represented an evolution-time interval and re-derived the well-known position and momentum propagators for a free massive relativistic particle. It was noted that time inversion represents conjugate unitary evolution (i.e. reflection about the real axis), because the unitarity of the representation implies $\rho(g^{-1})=\rho(g)^{-1}=\rho(g)^\dag$ is a symmetry in the Mellin transform. Further, one can check by direct calculation that the relativistic propagators are invariant under translation $|\tau|\rightarrow|\tau|\pm 1$ with $\tau\in i\R_{\pm}$. Apparently, in a relativistic setting unitary Klein-Gordon propagation is invariant under both inversion and translation of the evolution-time interval. What about consecutive inversion/translation transformations? They're not generally the same; unless $m=0$. This is a significant hint about the nature of evolution generated by real one-parameter subgroups $\phi_{\,\mathfrak{h}_{\mathrm{U}}}(\R)$.

Moving up to $g\in\phi_{\,\mathfrak{h}_{\mathrm{U}}}(\C)$, the hint suggests we should consider $\tau\in\C^\times$.\footnote{Of course complex $\tau$ renders $e^{-H(\tau)}$ non-unitary. Nevertheless, conjugate-evolution-time still obeys $\tau^\dag=\tau^{-1}$ so the representation $\rho$ remains unitary and functional Mellin still provides a representation.} Then, as an evolution-time interval requires a bounded region in $i\R_\pm$; by analogy we should consider bounded regions in $\C^\times$ to obtain an evolution interpretation. In the real case, the topology of $\tau$ parametrizing evolution-time can only be a line segment, but in the complex case $\tau$ can conceivably be any bounded Riemann surface with two marked points. Here we refer to \cite[$\S$5.2]{LA2} which found a resemblance between the functional positive power of a positive-definite operator and tree-level tachyon string scattering --- both open and closed. A positive power operator characterizes a resolvent, and the associated one-parameter subgroup represents the spectrum of the operator (as opposed to its evolution-time). Hence, the localization in that reference  used $\phi_{\,\mathfrak{h}_{\mathrm{U}}}(\R)\rightarrow\R^\times$ where $\R^\times$ double covers $\R_+$, but in the present context we would use $\phi_{\,\mathfrak{h}_{\mathrm{U}}}(\C)\rightarrow\R^\times$ and interpret $\tau\in\R^\times$ as the conformal equivalent of the boundary on the punctured disk.

  Although $e^{-\mathrm{H}(g)}$ is not local in this formalism, after topological localization its representation $e^{-H(\tau)}$ \emph{is} local in $\tau$. If $\mathcal{M}_\Gamma[\mathrm{E}^{-\mathrm{H}};\alpha]$ happens to be conformally invariant for some $\alpha\in\mathbb{S}$, then $\langle\psi_i|\mathcal{M}_\Gamma[\mathrm{E}^{-\mathrm{H}};\alpha]|\psi_j\rangle
  =\mathcal{M}_\Gamma[\langle\psi_i|\mathrm{E}^{-\mathrm{H}}|\psi_j\rangle;\alpha]$ localizes to the generating functional for $n$-point functions of a $2$-D CFT on $\C^\times$ with four marked points; two from $\tau$ and two from $\psi_i,\,\psi_j$. Generalizing further, one can contemplate localizing a collection of complex one-parameter subgroups $\phi_{\,\mathfrak{h}^{(1)}_{\mathrm{U}}}(\C)\times\phi_{\,\mathfrak{h}^{(2)}_{\mathrm{U}}}(\C)\times\ldots
\rightarrow(\C^\times)^n$.  Of course each subgroup factor would carry its own complex time $\tau^{(i)}$. As an example that illustrates this idea, unitary evolution of $\lambda\phi^4$ theory on AdS space leads to an $n$-point function for a conformal field source $\mathcal{O}$ of dimension $\Delta$ on the conformal boundary $\Sigma$ that goes like\cite[pg. 72]{Kap}
\begin{eqnarray}
  \langle\mathcal{O}(\sigma_1)\mathcal{O}(\sigma_2)\cdots\mathcal{O}(\sigma_n)\rangle
  &\sim&\int_{AdS} \left(\prod_{i=1}^n\int_0^\infty e^{-H(\sigma_i,x)\tau_i}\,\tau_i^{\Delta}\;d\log(\tau_i)\right)dx\notag\\
  &=&\int_{AdS}\prod_i\langle\psi_{\sigma_i}|\mathcal{M}_\Gamma[
  \mathrm{E}^{-\mathrm{H}^{(i)}};\Delta]|\psi_{x}\rangle\;dx\notag\\
  &=&\int_{AdS}\prod_i\langle\psi_{\sigma_i}|(H^{(i)})^{-\Delta}_\Gamma|\psi_{x}\rangle\;dx\notag\\
  &=&\int_{AdS}\prod_i\Bold{K}_{H^{(i)}}(\sigma_i,x)\;dx\notag\\
  &=&\prod_i\mathrm{tr}_x\Bold{K}_{H^{(i)}}(\sigma_i,x)
\end{eqnarray}
where $\sigma_i\in\Sigma$ are projective-cone coordinates, the Hamiltonian $H(\sigma_i,x)=2\sigma_i\cdot x$, and $\Bold{K}_{H^{(i)}}(\sigma_i,x):=\langle\psi_{\sigma_i}|(H^{(i)})^{-\Delta}_\Gamma|\psi_{x}\rangle$ is the point-to-boundary kernel.

 Consider a given evolution operator $\mathit{\Pi}_\lambda^{(1)}(\mathrm{U}_\tau)$ applied to a state $\psi(\tau_a)$. If we specify a final state $\psi(\tau_b)$ and look at a ``classical'' (i.e. extremal) evolution \emph{path} from $\tau_a$ to $\tau_b$, off hand it appears $\Im(\tau)$ parametrizes a change in state while $\Re(\tau)$ parametrizes a change in entropy. Evidently the ``classical'' evolution looks like non-equilibrium evolution in this case. On the other hand, if the system is governed by conformal invariance this interpretation is ambiguous, because such a split is not available and there does not appear to be a preferred path. In this case, one must employ the full machinery of 2-D CFT, and evolution-time/entropy have no invariant meaning.

\end{document}